\documentclass[11pt]{article}

\usepackage{times}
\usepackage{natbib}
\bibpunct{(}{)}{;}{a}{,}{,}
\usepackage{fullpage}
\usepackage{graphicx}
\usepackage{amsmath,amssymb,amsthm}
\usepackage{bm}
\usepackage{latexsym}
\usepackage{hyphenat}
\usepackage[pdftex,colorlinks=true,linkcolor=blue,citecolor=blue,urlcolor=blue,bookmarks=false,pdfpagemode=None]{hyperref}
\usepackage{url}
\usepackage{verbatim}
\usepackage{fancyhdr}
\usepackage{setspace}
\usepackage{paralist}
\usepackage{boxedminipage}
\usepackage{rotating}

\newcommand{\bv}{\begin{array}}
\newcommand{\ev}{\end{array}}
\newcommand{\bit}{\begin{itemize}}
\newcommand{\eit}{\end{itemize}}
\newcommand{\ben}{\begin{enumerate}}
\newcommand{\een}{\end{enumerate}}
\newcommand{\beq}{\begin{equation}}
\newcommand{\eeq}{\end{equation}}
\newcommand{\bvq}{\begin{eqnarray}}
\newcommand{\evq}{\end{eqnarray}}
\newcommand{\myfig}[1]{Figure~\ref{fig:#1}}
\newcommand{\mysec}[1]{Section~\ref{sec:#1}}
\newcommand{\myeq}[1]{Equation~\ref{eq:#1}}
\newcommand{\E}{\mathbb{E}}
\newcommand{\g}{\, | \,}
\newcommand{\ptoq}{p \rightarrow q}
\newcommand{\pfromq}{p \leftarrow q}

\begin{document}
\thispagestyle{empty}
\section*{\centerline{\LARGE Mixed Membership Stochastic Blockmodels}\vspace{20pt}\newline 
          \it \normalsize 
          Edoardo M. Airoldi, Princeton University \hfill (eairoldi@princeton.edu)\newline
          David M. Blei, Princeton University \hfill (blei@cs.princeton.edu)\newline
          Stephen E. Fienberg, Carnegie Mellon University \hfill (fienberg@stat.cmu.edu)\newline
          Eric P. Xing, Carnegie Mellon University \hfill (epxing@cs.cmu.edu)}

\begin{abstract}
 Observations consisting of measurements on relationships for pairs of objects arise in many settings, such as protein interaction and gene regulatory networks, collections of author-recipient email, and social networks. Analyzing such data with probabilisic models can be delicate because the simple exchangeability assumptions underlying many boilerplate models no longer hold.  
 In this paper, we describe a latent variable model of such data called the \textit{mixed membership stochastic blockmodel}.
 This model extends blockmodels for relational data to ones which capture mixed membership latent relational structure, thus providing an object-specific low-dimensional representation. 
 We develop a general variational inference algorithm for fast approximate posterior inference.  We explore applications to social and protein interaction networks.\newline

\noindent\textbf{Keywords:}  Hierarchical Bayes, Latent Variables, Mean-Field Approximation, Statistical Network Analysis, Social Networks, Protein Interaction Networks. 
\end{abstract}

\section{Introduction}
\label{sec:introduction}

 Modeling relational information among objects, such as pairwise relations represented as graphs, is becoming an important problem in modern data analysis and machine learning. Many data sets contain
interrelated observations. For example, scientific literature connects papers by citation, the Web connects pages by links, and protein-protein interaction data connects proteins by physical
interaction records. In these settings, we often wish to infer hidden attributes of the objects from the observed measurements on pairwise properties.  For example, we might want to
compute a clustering of the web-pages, predict the functions of a protein, or assess the degree of relevance of a scientific abstract to a scholar's query.

Unlike traditional attribute data collected over individual objects,
 \textit{relational data} violate
the classical independence or exchangeability assumptions that are
typically made in machine learning and statistics.  In fact, the
observations are interdependent by their very nature, and this interdependence
necessitates developing special-purpose statistical machinery for
analysis.

There is a history of research devoted to this end.  One problem
that has been heavily studied is that of \textit{clustering} the
objects to uncover a group structure based on the observed patterns
of interactions.  Standard model-based clustering methods, e.g.,
mixture models, are not immediately applicable to relational data
because they assume that the objects are conditionally independent
given their cluster assignments.  The latent stochastic
blockmodel~\citep{Snij:Nowi:1997} represents an adaptation of
mixture modeling to dyadic data.  In that model, each object belongs
to a cluster and the relationships between objects are governed by
the corresponding pair of clusters.  Via posterior inference on such
a model one can identify latent roles that objects possibly play,
which govern their relationships with each other. This model
originates from the stochastic blockmodel, where the roles of
objects are known in advance~\citep{Wang:Wong:1987}. A recent
extension of this model relaxed the finite-cardinality assumption on
the latent clusters, via a nonparametric hierarchical Bayesian
formalism based on the Dirichlet process prior~\citep{Kemp:Grif:Tene:2004,Kemp:Tene:Grif:etal:2006}.

 The latent stochastic blockmodel suffers from a limitation
that each object can only belong to one cluster, or in other words,
play a single latent role. In real life, it is not uncommon
to encounter more intriguing data on entities that are multi-facet.
For example, when a protein or a social actor interacts with
different partners, different functional or social contexts may
apply and thus the protein or the actor may be acting according to
different latent roles they can possible play. In this paper, we
relax the assumption of single-latent-role for actors, and develop a
\textit{mixed membership model} for relational data. Mixed
membership models, such as latent Dirichlet
allocation~\citep{Blei:Ng:Jord:2003}, have emerged in recent years
as a flexible modeling tool for data where the single cluster
assumption is violated by the heterogeneity within of a data point.
They have been successfully applied in many domains, such as
document analysis~\citep{Mink:Laff:2002,Blei:Ng:Jord:2003,Bunt:Jaku:2006},
surveys~\citep{Berk:Sing:Mant:1989,Eros:2002a}, image
processing~\citep{Fei-Fei:2005}, transcriptional regulation
\citep{Airo:Fien:Xing:2006}, and population genetics~\citep{Prit:Step:Rose:Donn:2000}.


 The mixed membership model associates each unit of observation with multiple clusters rather than a single cluster, via a membership probability-like vector.  The concurrent
membership of a data in different clusters can capture its different
aspects, such as different underlying topics for words constituting
each document.  The mixed membership formalism
is a particularly natural idea for relational data, where the
objects can bear multiple latent roles or cluster-memberships that
influence their relationships to others. As we will demonstrate, a
mixed membership approach to relational data lets us describe the
interaction between objects playing multiple roles. For example,
some of a protein's interactions may be governed by one function;
other interactions may be governed by another function.

 Existing mixed membership models are not appropriate for
relational data because they assume that the data are conditionally
independent given their latent membership vectors.  In relational
data, where each object is described by its relationships to others,
we would like to assume that the ensemble of mixed membership vectors
help govern the relationships of each object.  The conditional
independence assumptions of modern mixed membership models do not
apply.

In this paper, we develop mixed membership models for relational data,
develop a fast variational inference algorithm for inference and
estimation, and demonstrate the application of our technique to large
scale protein interaction networks and social networks.  Our model
captures the multiple roles that objects exhibit in interaction with
others, and the relationships between those roles in determining the
observed interaction matrix.

 Mixed membership and the latent block structure can be reliably recovered from relational data (Section \ref{sec:simulations}). 
 The application to a friendship network among students tests the model on a real data set where a well-defined latent block structure exists (Section \ref{sec:adolescent_data}).
 The application to a protein interaction network tests to what extent our model can reduce the dimensionality of the data, while revealing substantive information about the functionality of proteins that can be used to inform subsequent analyses (Section \ref{sec:protein_data}).

\section{The mixed membership stochastic blockmodel}
\label{sec:alb}

\begin{figure}[t!]
  \centering
  \includegraphics[width=13cm]{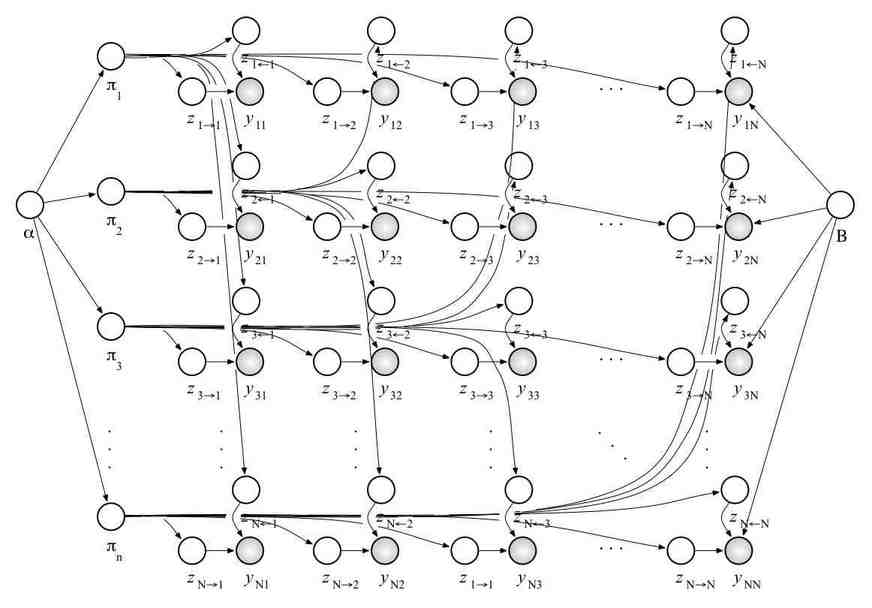}
  \caption{A graphical model of the mixed membership stochastic
    blockmodel.  We did not draw all the arrows out of the block model
    $B$ for clarity.  All the interactions $R(p,q)$ depend on it.}
  \label{fig:alb}
\end{figure}

In this section, we describe the modeling assumptions behind our
proposed mixed membership model of relational data.  We represent
observed relational data as a graph $G=({\cal N},R)$, where $R(p,q)$ maps
pairs of nodes to values, i.e., edge weights.  In this work, we
consider binary matrices, where $R(p,q) \in \{0,1\}$.  Thus, the data
can be thought of as a directed graph.

As a running example, we will reanalyze the monk data of
\citet{Samp:1968}.  Sampson measured a collection of sociometric
relations among a group of monks by repeatedly asking questions such
as ``Do you like X?'' or ``Do you trust X?'' to determine asymmetric
social relationships within the group.  The questionnaire was repeated
at four subsequent epochs.  Information about these repeated,
asymmetric relations can be collapsed into a square binary table that
encodes the directed connections between
monks~\citep{Brei:Boor:Arab:1975}.  In analyzing this data, the goal
is to determine the underlying social groups within the monastary.

In the context of the monastery example, we assume $K$ latent groups
over actors, and the observed network is generated according to
latent distributions of group-membership for each monk and a matrix
of group-group interaction strength.  The latent per-monk
distributions are specified by simplicial vectors. Each monk is
associated with a randomly drawn vector, say $\vec{\pi}_i$ for monk
$i$, where $\pi_{i,g}$ denotes the probability of monk $i$ belonging
to group $g$. That is, each monk can simultaneously belong to
multiple groups with different degrees of affiliation strength. The
probabilities of interactions between different groups are defined
by a matrix of Bernoulli rates $B_{(K\times K)}$, where $B(g,h)$
represents the probability of having a link between a monk from
group $g$ and a monk from group $h$.

For each network node (i.e., monk), the indicator vector
$\vec{z}_{p\rightarrow q}$ denotes the group membership of node $p$
when it is approached by node $q$ and $\vec{z}_{p\leftarrow q}$
denotes the group membership of node $q$ when it is approached by
node $p$~\footnote{An indicator vector is used to denote membership in one of the $K$ groups. Such a membership-indicator vector is specified as a $K$-dimensional
  vector of which only one element equals to one, whose index corresponds to the group to be indicated, and all other elements equal to zero.}.  $N$ denotes the number of nodes in the graph, and
recall that $K$ denotes the number of distinct groups a node can
belong to. Now putting everything together, we have a mixed
membership stochastic blockmodel (MMSB), which posits that a graph
$G=({\cal N},R)$ is drawn from the following procedure.
\begin{itemize}
\item For each node $p \in {\cal N}$:
  \begin{itemize}
  \item Draw a $K$ dimensional mixed membership vector $\vec{\pi}_p \sim
    \textrm{Dirichlet}\bigm( \vec\alpha \bigm)$.
  \end{itemize}
\item For each pair of nodes $(p,q) \in {\cal N} \times {\cal N}$:
  \begin{itemize}
  \item Draw membership indicator for the initiator, $\vec z_{p
      \rightarrow q} ~ \sim {\rm Multinomial}\bigm(\vec\pi_{p} \bigm)$.
  \item Draw membership indicator for the receiver, $\vec z_{q
      \rightarrow p} ~ \sim {\rm Multinomial}\bigm(\vec\pi_{q} \bigm)$.
  \item Sample the value of their interaction, $R(p,q) \sim {\rm
      Bernoulli}\bigm(\vec{z}_{\ptoq}^{~\top} B ~ \vec{z}_{\pfromq}
    \bigm)$.
  \end{itemize}
\end{itemize}
This process is illustrated as a graphical model in \myfig{alb}.  Note
that the group membership of each node is {\em context dependent}.
That is, each node may assume different membership when interacting to
or being interacted by different peers.  Statistically, each node is
an admixture of group-specific interactions.  The two
sets of latent group indicators are denoted by $\{ \vec{z}_{\ptoq} :
p,q \in \mathcal{N} \} =: Z_\rightarrow$ and $\{ \vec{z}_{\pfromq} :
p,q \in \mathcal{N} \} =: Z_\leftarrow$. 
Also note that the pairs of group memberships that underlie interactions, for example, $(\vec{z}_{\ptoq},\vec{z}_{\pfromq})$ for $R(p,q)$, need not be equal; this fact is useful for characterizing asymmetric interaction networks. Equality may be enforced  when modeling symmetric interactions.

Under the MMSB, the joint probability of the data $R$ and the latent
variables $\{\vec{\pi}_{1:N},Z_\rightarrow,Z_\leftarrow\}$ can be
written in the following factored form,
\begin{eqnarray}
  \label{eq:likelihood_with_zs}
  \lefteqn{p ( R,\vec{\pi}_{1:N},Z_\rightarrow,Z_\leftarrow |
    \vec{\alpha},B ) }  \nonumber \\ &&
  =
  \prod_{p,q} ~ P(R(p,q) | \vec{z}_{\ptoq}, \vec{z}_{\pfromq}, B)
  P(\vec{z}_{\ptoq} | \vec{\pi}_p) P(\vec{z}_{\pfromq} | \vec{\pi}_q)
  \prod_p P(\vec{\pi}_p | \vec{\alpha}).
\end{eqnarray}
 This model easily generalizes to two important cases.
(Appendix \ref{app:general_formulation} develops this intuition more formally.)
First, multiple networks among the same actors can be generated by the same latent vectors.  This might be useful, for example, for analyzing simultaneously the relational measurements about esteem and disesteem, liking and disliking, positive influence and negative influence, praise and blame, e.g., see \cite{Samp:1968}, or those about the collection of 17 relations measured by \cite{Brad:1987}.
Second, in the MMSB the data generating distribution is a Bernoulli, but $B$ can be a matrix that parameterizes any kind of distribution.  For example, technologies for measuring interactions between pairs of proteins such as mass spectrometry \citep{Ho:Gruh:Heil:etal:2002} and tandem affinity purification \citep{Gavi:Bosc:Krau:etal:2002} return a probabilistic assessment about the presence of interactions, thus setting the range of $R(p,q)$ to $[0,1]$. This is not the case for the manually curated collection of interactions we analyze in Section \ref{sec:protein_data}.

 The central computational problem for this model is  computing the posterior distribution of per-node mixed
membership vectors and per-pair roles that generated the data.  The
membership vectors in particular provide a low dimensional
representation of the underlying objects in the matrix of
observations, which can be used in the data analysis task at hand.  A
related problem is parameter estimation, which is to find maximum
likelihood estimates of the Dirichlet parameters $\vec{\alpha}$ and
Bernoulli rates $B$.

For both of these tasks, we need to compute the probability of the
observed data.  This amounts to marginalizing out the latent variables
from \myeq{likelihood_with_zs}.  This is intractable for even small
graphs.  In \mysec{estimation_inference}, we develop a fast
variational algorithm to approximate this marginal likelihood for
parameter estimation and posterior inference.

\subsection{Modeling sparsity}
\label{sec:sparsity}

Many real-world networks are sparse, meaning that most pairs of
nodes do not have edges connecting them.  For many pairs, the
absence of an interaction is a result of the rarity of any
interaction, rather than an indication that the underlying latent
groups of the objects do not tend to interact.  In the MMSB,
however, all observations (both interactions and non-interactions)
\textit{contribute equally} to our inferences about group memberships and group to group interaction patterns. It is thus
useful, in practical applications, to account for sparsity.

We introduce a sparsity parameter $\rho \in [0,1]$ to calibrate the
importance of non-interaction.  This models how often a
non-interaction is due to sparsity rather than carrying information
about the group memberships of the nodes.  Specifically, instead of
drawing an edge directly from the Bernoulli specified above, we
downweight it to $(1-\rho) \cdot \vec{z}_{\ptoq}^{~\top} B ~
\vec{z}_{\pfromq}$.  The probability of having no interaction is thus
$1-\sigma_{pq} = (1-\rho) \cdot \vec{z}_{\ptoq}^{~\top} (1-B) ~
\vec{z}_{\pfromq} + \rho$.  (This is equivalent to re-parameterizing
the interaction matrix $B$.)  In posterior inference and parameter
estimation, a large value of $\rho$ will cause the interactions in the
matrix to be weighted more than non-interactions in determining the
estimates of $\{\vec{\alpha},B,\vec{\pi}_{1:N}\}$.

\subsection{A case study of the Monastery network via MMSB: crisis in a Cloister}
\label{sec:monk_data}

\begin{figure}[b!]
  \centering
  \includegraphics[width=0.40\textwidth]{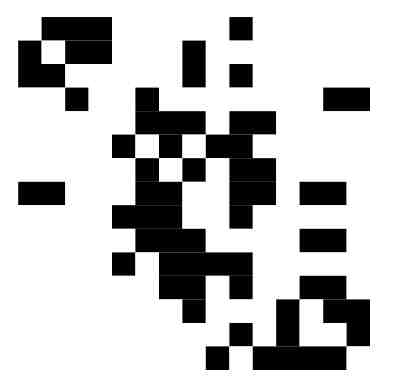}\hfill
  \includegraphics[width=0.40\textwidth]{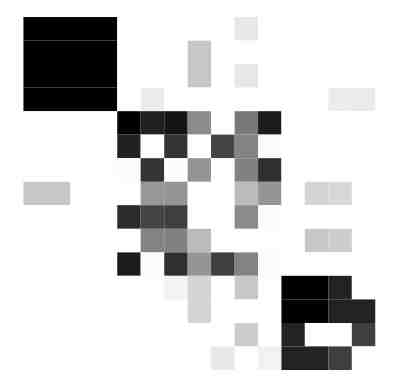}
  \caption{Original matrix of sociometric relations (left), and
    estimated relations obtained by the model (right).}
 \label{fig:data_exp}
\end{figure}

Before turning to the details of posterior inference and parameter
estimation, we illustrate the MMSB with an analysis of the monk data
described above.  In more detail, \cite{Samp:1968} surveyed 18 novice
monks in a monastery and asked them to rank the other novices in terms
of four sociometric relations: like/dislike, esteem, personal
influence, and alignment with the monastic credo.  We consider
Breiger's collation of Sampson's data \citep{Brei:Boor:Arab:1975}.
The original graph of monk-monk interaction is illustrated in
\myfig{data_exp} (left).

Sampson spent several months in a monastery in New England, where
novices (the monks) were preparing to join a monastic order.
Sampson's original analysis was rooted in direct anthropological
observations. He strongly suggested the existence of tight factions
among the novices: the loyal opposition (whose members joined the
monastery first), the young turks (who joined later on), the
outcasts (who were not accepted in the two main factions), and the
waverers (who did not take sides). The events that took place during
Sampson's stay at the monastery supported his observations---members
of the young turks resigned after their leaders were expelled over religious
differences (John Bosco and Gregory). We shall refer to the labels
assigned by Sampson to the novices in the analysis below. For more
analyses, we refer to \cite{Fien:Meye:Wass:1985}, \cite{Davi:Carl:2006} and
\cite{Hand:Raft:Tant:2006}.

Using the techniques outlined below in \mysec{estimation_inference},
we fit the monks to MMSB models for different numbers of groups,
providing model estimates $\{\hat{\alpha}, \hat{B}\}$ and posterior
mixed membership vectors $\vec{\pi}_n$ for each monk.
Here, we use the following approximation to BIC to choose the number of groups in the MMSB:
\[
  BIC = 2 \cdot \log p(R) \approx 2 \cdot \log p(R |
  \widehat{\vec{\pi}}, \widehat{Z}, \widehat{\vec{\alpha}},
  \widehat{B}) - |\vec{\alpha},B| \cdot \log |R|,
\]
 which selects three groups, where $|\vec{\alpha},B|$ is the number of hyper-parameters in the model, and $|R|$ is the number of positive relations observed~\citep{Voli:Raft:2000,Hand:Raft:Tant:2006}. 
 Note that this is the same number of groups that Sampson identified.  We illustrate the fit of model fit via the predicted network in \myfig{data_exp} (Right).

\begin{figure}[t]
 \centering
  \includegraphics[width=\textwidth]{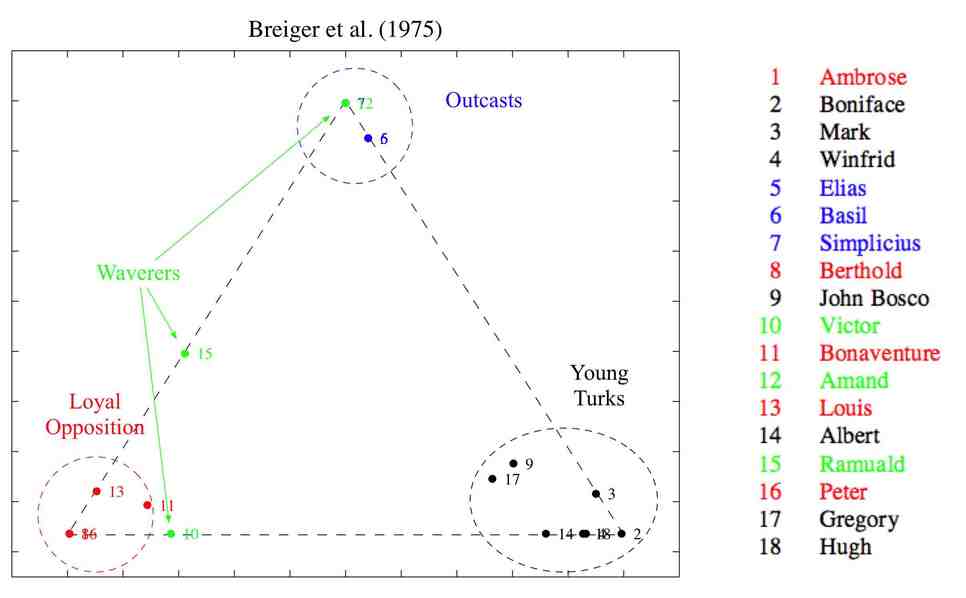}
  \caption{Posterior mixed membership vectors, $\vec{\pi}_{1:18}$, projected in the simplex. The estimates correspond to a model with $B := \mathbb{I}_3$, and $\hat \alpha = 0.058$. Numbered points can be mapped to monks' names using the legend on the right. The colors identify the four factions defined by Sampson's anthropological observations.}
 \label{fig:simplex}
\end{figure}
\begin{figure}
 \centering
  \includegraphics[width=0.75\textwidth]{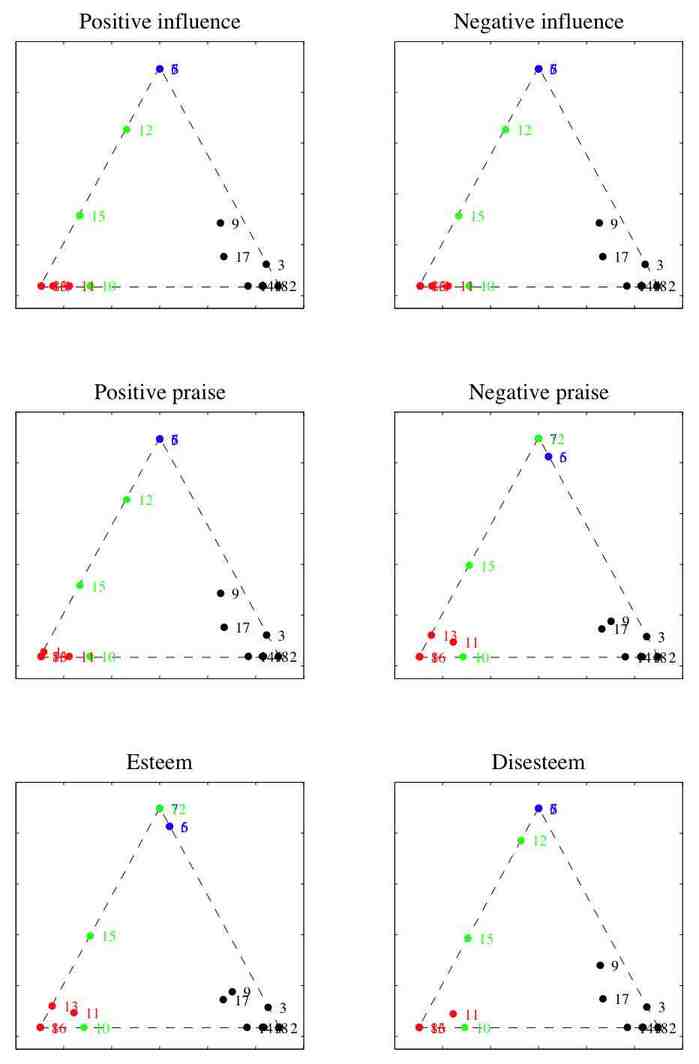}
  \caption{Independent analyses of six graphs encoding the relations: positive and negative influence, positive and negative praise, esteem and disesteem. Posterior mixed membership vectors for each graph, corresponding to models with $B := \mathbb{I}_3$, and $\hat \alpha$ via empirical Bayes, are projected in the simplex. (Legend as in Figure \ref{fig:simplex}.)}
 \label{fig:simplex_2}
\end{figure}

The MMSB can provide interesting descriptive statistics about the
actors in the observed graph.  In \myfig{simplex} we illustrate the
the posterior means of the mixed membership scores, $\mathbb{E} [
\vec{\pi} | R ]$, for the 18 monks in the monastery.  Note that the
monks cluster according to Sampson's classification, with Young Turks,
Loyal Opposition, and Outcasts dominating each corner respectively.
We can see the central role played by John Bosco and Gregory, who exhibit
relations in all three groups, as well as the uncertain affiliations
of Ramuald and Victor; Amand's uncertain affiliation, however, is not captured.

 Later, we considered six graphs encoding specific relations---positive and negative influence, positive and negative praise, esteem and disesteem---and we performed independent analyses using MMSB. This allowed us to look for signal about the mixed membership of monks to factions that may have been lost in the data set prepared by \citet{Brei:Boor:Arab:1975} because of averaging.
 Figure \ref{fig:simplex_2} shows the projections in the simplex of the posterior mixed membership vectors for each of the six relations above. For instance, we can see how Victor, Amand, and Ramuald---the three waverers---display mixed membership in terms of positive and negative influence, positive praise, and disesteem. The mixed membership of Amand, in particular, is expressed in terms of these relations, but not in terms of negative praise or esteem. This finding is supported Sampson's anthropological observations, and it suggests that relevant substantive information has been lost when the graphs corresponding to multiple sociometric relations have been collapsed into a single social network \citep{Brei:Boor:Arab:1975}.
 Methods for the analysis of multiple sociometric relations are thus to be preferred. In Appendices \ref{app:general_formulation} and \ref{app:general_inference} extend the mixed membership stochastic blockmodel to deal with the case of multivariate relations, and we solve estimation and inference in the general case.

\section{Parameter Estimation and Posterior Inference}
\label{sec:estimation_inference}

In this section, we tackle the two computational problems for the
MMSB: posterior inference of the per-node mixed membership vectors and
per-pair roles, and parameter estimation of the Dirichlet parameters
and Bernoulli rate matrix.  We use empirical Bayes to estimate the
parameters $(\vec{\alpha},B)$, and employ a mean-field approximation
scheme \citep{Jord:Ghah:Jaak:Saul:1999} for posterior inference.

\subsection{Posterior inference}

In posterior inference, we would like to compute the posterior
distribution of the latent variables given a collection of
observations.  As for other mixed membership models, this is
intractable to compute.  The normalizing constant of the posterior
is the marginal probability of the data, which requires an
intractable integral over the simplicial vectors $\vec{\pi}_p$,
\begin{equation}
  \label{eq:marginal}
  p ( R | \vec{\alpha},B ) = \int_{\vec{\pi}_{1:N}}
  \prod_{p,q} \sum_{z_{\pfromq}, z_{\ptoq}} ~ P(R(p,q) | \vec{z}_{\ptoq}, \vec{z}_{\pfromq}, B)
  P(\vec{z}_{\ptoq} | \vec{\pi}_p) P(\vec{z}_{\pfromq} | \vec{\pi}_q)
  \prod_p P(\vec{\pi}_p | \vec{\alpha}).
\end{equation}

A number of approxiate inference algorithms for mixed membership
models have appeared in recent years, including mean-field variational
methods \citep{Blei:Ng:Jord:2003,Teh:Newm:Well:2007}, expectation
propagation \citep{Mink:Laff:2002}, and Monte Carlo Markov chain
sampling (MCMC) \citep{Eros:Fien:2005,Grif:Stey:2004}.

We appeal to mean-field variational methods to approximate the
posterior of interest.  Mean-field variational methods provide a
practical deterministic alternative to MCMC.  MCMC is not practical
for the MMSB due to the large number of latent variables needed to be
sampled.  The main idea behind variational methods is to posit a
simple distribution of the latent variables with free parameters.
These parameters are fit to be close in Kullback-Leibler divergence to
the true posterior of interest.  Good reviews of variational methods
method can be found in a number of papers~\citep{Jord:Ghah:Jaak:Saul:1999,Wain:Jord:2003b,Xing:Jord:Russ:2003,Bish;Spie;Winn;2003}

The log of the marginal probability in \myeq{marginal} can be bound
with Jensen's inequality as follows,
\begin{equation}
\label{eq:bound}
  \log p(R \g \alpha, B) \geq \E_q \bigm[ \log p(R, \vec{\pi}_{1:N},
  Z_{\rightarrow}, Z_{\leftarrow}
  | \alpha, B)\bigm] - \E_q \bigm[ \log q(\vec{\pi}_{1:N},
  Z_{\rightarrow}, Z_{\leftarrow}) \bigm],
\end{equation}
by introducing a distribution of the latent variables $q$ that depends on a set of free parameters
We specify $q$ as the mean-field fully-factorized family,
\begin{equation}
  q(\vec{\pi}_{1:N},Z_\rightarrow,Z_\leftarrow |
  \vec{\gamma}_{1:N},\Phi_\rightarrow,\Phi_\leftarrow) = \prod_{p} ~ q_1 (\vec{\pi}_p | \vec{\gamma}_p) ~ \prod_{p,q} ~ \Big( q_2 (\vec{z}_{\ptoq} | \vec{\phi}_{\ptoq}) ~ q_2
(\vec{z}_{\pfromq} | \vec{\phi}_{\pfromq}) \Big),
\end{equation}
where $q_1$ is a Dirichlet, $q_2$ is a multinomial, and
$\{\vec{\gamma}_{1:N},\Phi_\rightarrow,\Phi_\leftarrow\}$ are the set
of free {\em variational parameters} that can be set to tighten the
bound.

Tightening the bound with respect to the variational parameters is
equivalent to minimizing the KL divergence between $q$ and the true
posterior.  When all the nodes in the graphical model are conjugate
pairs or mixtures of conjugate pairs, we can directly write down a
coordinate ascent algorithm for this
optimization~\citep{Xing:Jord:Russ:2003,Bish;Spie;Winn;2003}.  The
update for the variational multinomial parameters is
\begin{eqnarray}
  \label{eq:phi_to}
  \hat \phi_{{\ptoq, g}} &\propto& e^{~\mathbb{E}_q \bigm[ \log
    \pi_{p,g} \bigm]} \cdot \prod_h \biggm( B(g,h)^{R(p,q)} \cdot
  \bigm(1-B(g,h) \bigm)^{1-R(p,q)} \biggm)^{\phi_{\pfromq,h}} \\
  \label{eq:phi_from}
   \hat \phi_{{\pfromq, h}}
   &\propto&  e^{~\mathbb{E}_q \bigm[ \log \pi_{q,h} \bigm]} \cdot
 \prod_g \biggm( B(g,h)^{R(p,q)} \cdot \bigm(1-B(g,h)
 \bigm)^{1-R(p,q)} \biggm)^{\phi_{\ptoq,g}},
\end{eqnarray}
for $g,h=1, \dots, K$.  The update for the variational Dirichlet
parameters $\gamma_{p,k}$ is
\begin{equation}
  \label{eq:gamma}
  \hat \gamma_{p,k} = \alpha_k + \sum_{q} \phi_{\ptoq,k} + \sum_{q} \phi_{\pfromq,k},
\end{equation}
for all nodes $p=1,\dots,N$ and $k=1, \dots, K$. An analytical expression for $\mathbb{E}_q \bigm[ \log \pi_{q,h} \bigm]$ is derived in Appendix \ref{app:exp_log_pi}. The complete coordinate ascent algorithm to perform variational inference is described in Figure \ref{fig:algs}.
\begin{figure}[t!]
\begin{center}
\begin{tabular}{lllll}
\\ \hline \\
 1. & \multicolumn{4}{l}{initialize $\vec{\gamma}_{pk}^0 = \frac{2N}{K}$ for all $p,k$} \\
 2. & \multicolumn{4}{l}{\textbf{repeat}} \\
 3. &~~& \multicolumn{3}{l}{\textbf{for} $p=1$ to $N$} \\
 4. &~~&~~& \multicolumn{2}{l}{\textbf{for} $q=1$ to $N$} \\
 5. &~~&~~&~~& get \textbf{variational} $\vec{\phi}_{\ptoq}^{t+1}$ and $\vec{\phi}_{\pfromq}^{t+1} = f \bigm( R(p,q), \vec{\gamma}_p^t, \vec{\gamma}_q^t, B^t \bigm)$ \\
 6. &~~&~~&~~& partially update $\vec{\gamma_{p}^{t+1}}$, $\vec{\gamma_{q}^{t+1}}$ and $B^{t+1}$ \\
 7. & \multicolumn{4}{l}{\textbf{until} convergence} \\ \\ \hline
\end{tabular}
\begin{tabular}{llll}
\\ \hline \\
 5.1. &\multicolumn{3}{l}{initialize $\phi_{\ptoq,g}^0 = \phi_{\pfromq,h}^0 = \frac{1}{K}$ for all $g,h$} \\
 5.2. &\multicolumn{3}{l}{\textbf{repeat}} \\
 5.3. &~~ &\multicolumn{2}{l}{\textbf{for} $g=1$ to $K$} \\
 5.4. &~~ &~~ & update $\phi_{\ptoq}^{s+1} \propto f_1 \bigm(\vec{\phi}_{\pfromq}^{s}, \vec{\gamma}_p, B \bigm)$ \\
 5.5. &~~ &\multicolumn{2}{l}{normalize $\vec{\phi}_{\ptoq}^{s+1}$ to sum to 1} \\
 5.6. &~~ &\multicolumn{2}{l}{\textbf{for} $h=1$ to $K$} \\
 5.7. &~~ &~~ & update $\phi_{\pfromq}^{s+1} \propto f_2 \bigm(\vec{\phi}_{\ptoq}^{s}, \vec{\gamma}_q, B \bigm)$ \\
 5.8. &~~ &\multicolumn{2}{l}{normalize $\vec{\phi}_{\pfromq}^{s+1}$ to sum to 1} \\
 5.9. &\multicolumn{3}{l}{\textbf{until} convergence} \\ \\ \hline \\
\end{tabular}
\end{center}
\caption{\textbf{Top:} The two-layered variational inference for
  $(\vec{\gamma},\phi_{\ptoq,g},\phi_{\pfromq,h})$ and $M=1$.  The
  inner algorithm consists of Step 5. The function $f$ is described in
  details in the bottom panel. The partial updates in Step 6 for
  $\vec{\gamma}$ and $B$ refer to Equation \ref{eq:gamma} of Section
  \ref{sec:inf_e} and Equation \ref{eq:b} of Section \ref{sec:inf_m},
  respectively.  \textbf{Bottom:} Inference for the variational
  parameters $(\vec{\phi}_{\ptoq}, \vec{\phi}_{\pfromq})$
  corresponding to the basic observation $R(p,q)$. This nested
  algorithm details Step 5 in the top panel. The functions $f_1$ and
  $f_2$ are the updates for $\phi_{\ptoq,g}$ and $\phi_{\pfromq,h}$
  described in Equations \ref{eq:phi_to} and \ref{eq:phi_from} of
  Section \ref{sec:inf_e}.}
\label{fig:algs}
\end{figure}

To improve convergence in the relational data setting, we introduce a {\em nested} variational inference
scheme based on an alternative schedule of updates to the traditional
ordering.  In a na\"ive iteration scheme for variational inference,
one initializes the variational Dirichlet parameters
$\vec{\gamma}_{1:N}$ and the variational multinomial parameters
$(\vec{\phi}_{\ptoq},\vec{\phi}_{\pfromq})$ to non\hyp{}informative
values, and then iterates until convergence the following two steps:
(i) update $\vec{\phi}_{\ptoq}$ and $\phi_{\pfromq}$ for all edges
$(p,q)$, and (ii) update $\vec{\gamma}_p$ for all nodes $p \in
\mathcal{N}$.  In such algorithm, at each variational inference cycle
we need to allocate $NK+2N^2K$ scalars.

In our experiments, the na\"ive variational algorithm often failed to
converge, or converged only after many iterations.  We attribute this
behavior to the dependence between $\vec{\gamma}_{1:N}$ and $B$, which
is not satisfied by the na\"ive algorithm.  Some intuition about why
this may happen follows.  From a purely algorithmic perspective, the
na\"ive variational EM algorithm instantiates a large coordinate
ascent algorithm, where the parameters can be semantically divided
into coherent blocks.  Blocks are processed in a specific order, and
the parameters within each block get all updated each
time.\footnote{Within a block, the order according to which (scalar)
  parameters get updated is not expected to affect convergence.}  At
every new iteration the na\"ive algorithm sets all the elements of
$\vec{\gamma}_{1:N}^{t+1}$ equal to the same constant.  This dampens
the likelihood by suddenly breaking the dependence between the
estimates of parameters in $\widehat{\vec{\gamma}}_{1:N}^t$ and in
$\hat B^t$ that was being inferred from the data during the previous
iteration.

Instead, the nested variational inference algorithm maintains some of
this dependence that is being inferred from the data across the
various iterations.  This is achieved mainly through a different
scheduling of the parameter updates in the various blocks.  To a minor
extent, the dependence is maintained by always keeping the block of
free parameters, $(\vec{\phi}_{\ptoq},\vec{\phi}_{\pfromq})$,
optimized given the other variational parameters.  Note that these
parameters are involved in the updates of parameters in
$\vec{\gamma}_{1:N}$ and in $B$, thus providing us with a channel to
maintain some of the dependence among them, i.e., by keeping them at
their optimal value given the data.

Furthermore, the nested algorithm has the advantage that it trades
time for space thus allowing us to deal with large graphs; at each
variational cycle we need to allocate $NK+2K$ scalars only.  The
increased running time is partially offset by the fact that the
algorithm can be parallelized and leads to empirically observed
faster convergence rates.  This algorithm is also better than MCMC
variations (i.e., blocked and collapsed Gibbs samplers) in terms of
memory requirements and convergence rates---an empirical comparison between the na\"ive and nested variational inference schemes in presented in Figure \ref{fig:simdata1}, left panel.

\subsection{Parameter estimation}

We compute the empirical Bayes estimates of the model hyper-parameters
$\{\vec{\alpha}, B\}$ with a variational expectation-maximization
(EM) algorithm.
 Alternatives to empirical Bayes have been proposed to fix the hyper-parameters and reduce the computation. The results, however, are not always satisfactory and often times cause of concern, since the inference is sensitive to the choice of the hyper-parameters \citep{Airo:Fien:Jout:Love:2006}. Empirical Bayes, on the other hand, guides the posterior inference towards a region of the hyper-parameter space that is supported by the data. 

 Variational EM uses the lower bound in \myeq{bound}
as a surrogate for the likelihood.  To find a local optimum of the
bound, we iterate between: fitting the variational distribution $q$
to approximate the posterior, and maximizing the corresponding lower bound for the likelihood 
with respect to the parameters.  The latter M-step is equivalent to
finding the MLE using expected sufficient statistics under the
variational distribution.
 We consider the maximization step for each parameter in turn. 
 
 A closed form solution for the approximate maximum likelihood estimate
of $\vec\alpha$ does not exist \citep{Mink:2003a}.  We use a
linear-time Newton-Raphson method, where the gradient and Hessian are
\begin{eqnarray*}
   \frac{\partial \mathcal{L}_{\vec{\alpha}}}{\partial \alpha_{k}} & = &
    N \biggm( \psi \bigm( \sum_{k} \alpha_{k} \bigm) - \psi (\alpha_{k}) \biggm) + \sum_{p} \biggm( \psi(\gamma_{p,k}) - \psi \bigm( \sum_{k} \gamma_{p,k} \bigm) \biggm),\\
   \frac{\partial \mathcal{L}_{\vec{\alpha}}}{\partial \alpha_{k_1} \alpha_{k_2}} & = &
    N \biggm( \mathbb{I}_{(k_1 = k_2)} \cdot \psi' (\alpha_{k_1}) - \psi' \bigm( \sum_{k} \alpha_{k} \bigm) \biggm).
\end{eqnarray*}
The approximate MLE of $B$ is \bvq
 \label{eq:b}
 \hat B (g,h) = \frac{\sum_{p,q} R(p,q) \cdot \phi_{\ptoq g} \,
   \phi_{\pfromq h}}{\sum_{p,q} \phi_{\ptoq g} \, \phi_{\pfromq h}},
 \evq for every index pair $(g,h) \in [1,K] \times [1,K]$.  Finally,
 the approximate MLE of the sparsity parameter $\rho$ is \bvq
\label{eq:rho}
\hat \rho = \frac{\sum_{p,q} \bigm( 1-R(p,q) \bigm) \cdot \bigm(
  \sum_{g,h} \phi_{\ptoq g} \, \phi_{\pfromq h}\bigm)}{\sum_{p,q}
  \sum_{g,h} \phi_{\ptoq g} \, \phi_{\pfromq h}}.
\evq

Alternatively, we can fix $\rho$ prior to the analysis; the density of
the interaction matrix is estimated with $\hat d = \sum_{p,q}
R(p,q)/N^2$, and the sparsity parameter is set to $\tilde \rho =
(1-\hat d)$.  This latter estimator attributes all the information in
the non-interactions to the point mass, i.e., to latent sources other
than the block model $B$ or the mixed membership vectors
$\vec{\pi}_{1:N}$.  It does however provide a quick recipe to reduce
the computational burden during exploratory analyses.\footnote{Note that
$\tilde \rho = \hat \rho$ in the case of single membership. In fact,
that implies $\phi_{\ptoq g}^m = \phi_{\pfromq h}^m =1$ for some
$(g,h)$ pair, for any $(p,q)$ pair.}

 Several model selection strategies are available for complex hierarchical models. In our setting, model selection translates into the determination of a plausible value of the number of groups $K$. In the various analyses presented, we selected the optimal value of $K$ with the highest averaged held-out likelihood in a cross-validation experiment, on large networks, and using an approximation to BIC, on small networks.

\section{Experiments and Results}

 Here, we present experiments on simulated data, and we develop two applications to social and protein interaction networks. The three problem settings serve different purposes.

 Simulations are performed in Section \ref{sec:simulations} to show that both mixed membership, $\vec{\pi}_{1:N}$, and the latent block structure, $B$, can be recovered from data, when they exist, and that the nested variational inference algorithm is as fast as the na\"ive implementation while reaching a higher peak in the likelihood---coeteris paribus.
 
 The application to a friendship network among students in Section \ref{sec:adolescent_data} tests the model on a real data set where we expect a well-defined latent block structure to inform the observed connectivity patterns in the network. In this application, the blocks are interpretable in terms of grades. We compare our results with those that were recently published with a simple mixture of blocks \citep{Dore:Bata:Ferl:2007} and with a latent space model \citep{Hand:Raft:Tant:2006} on the same data.
 
 The application to a protein interaction network in Section \ref{sec:protein_data} tests the model on a real data set where we expect a noisy, vague latent block structure to inform the observed connectivity patterns in the network to some degree. In this application, the blocks are interpretable in terms functional biological contexts. This application tests to what extent our model can reduce the dimensionality of the data, while revealing substantive information about the functionality of proteins that can be used to inform subsequent analyses.
  
\subsection{Exploring Expected Model Behavior with Simulations}
\label{sec:simulations}

 In developing the MMSB and the corresponding computation, our hope is
the the model can recover both the mixed membership of nodes to
clusters and the latent block structure among clusters in situations
where a block structure exists and the relations are measured with
some error.  To substantiate this claim, we sampled graphs of $100,
300,$ and $600$ nodes from blockmodels with $4, 10,$ and $20$
clusters, respectively, using the MMSB.  We used different values of
$\alpha$ to simulate a range of settings in terms of membership of
nodes to clusters---from almost unique $(\alpha=0.05)$ to mixed
$(\alpha=0.25)$.

\subsubsection{Recovering the Truth}

\begin{figure}[t!]
  \centering
    \includegraphics[width=4.5cm]{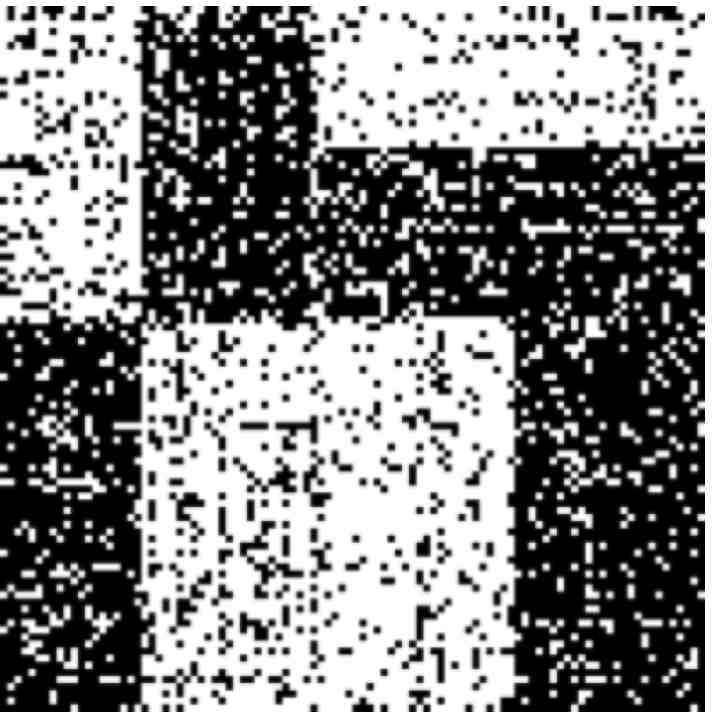}
    \includegraphics[width=4.5cm]{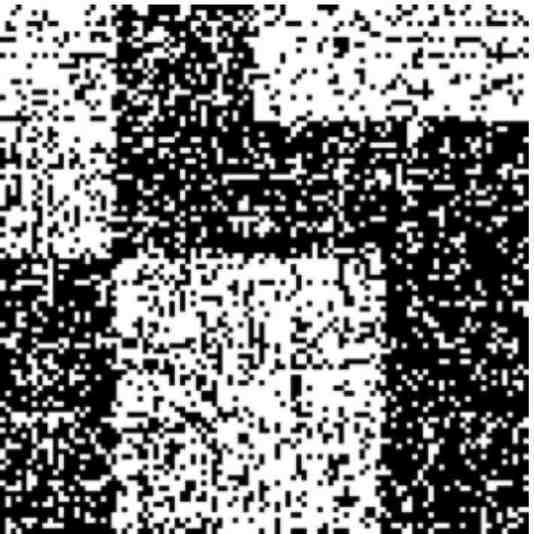}
    \includegraphics[width=4.5cm]{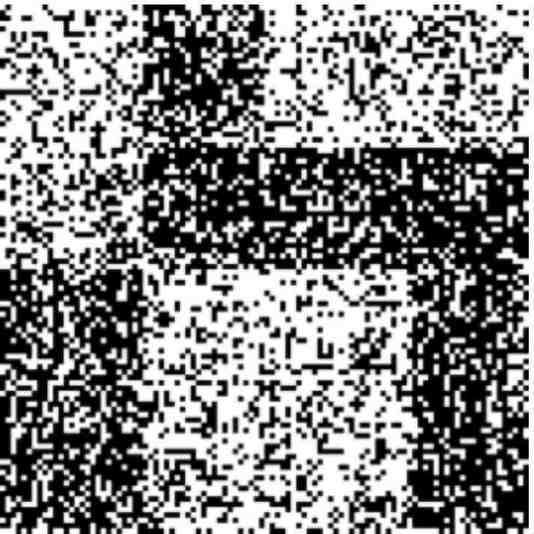}
  \centering
    \includegraphics[width=4.5cm]{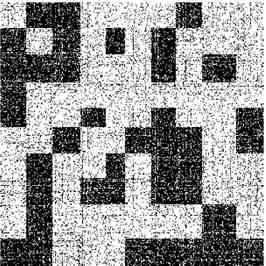}
    \includegraphics[width=4.5cm]{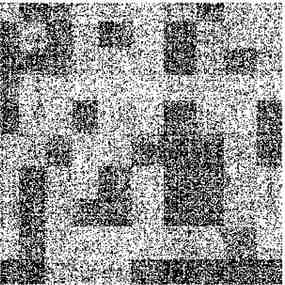}
    \includegraphics[width=4.5cm]{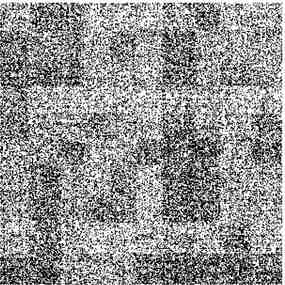}
  \centering
    \includegraphics[width=4.5cm]{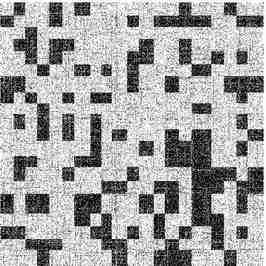}
    \includegraphics[width=4.5cm]{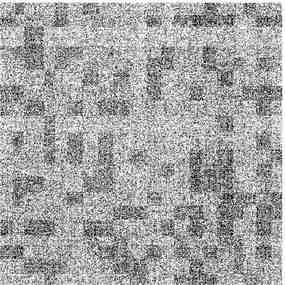}
    \includegraphics[width=4.5cm]{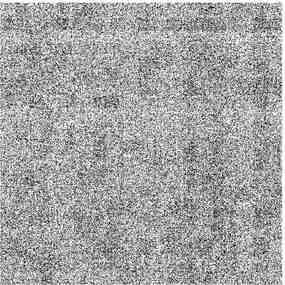}
    \caption{Adjacency matrices of corresponding to simulated interaction graphs with 100 nodes and 4 clusters, 300 nodes and 10 clusters, 600 nodes and 20 clusters (top to bottom) and $\alpha$ equal to $0.05, 0.1$ and $0.25$ (left to right).  Rows, which corresponds to nodes, are reordered according to their  most likely membership.  The estimated reordering reveals the original blockmodel in all the data settings we tested.}
 \label{fig:blk_mods}
\end{figure}

\begin{figure}[t!]
 \centering
 \includegraphics[width=7.75cm]{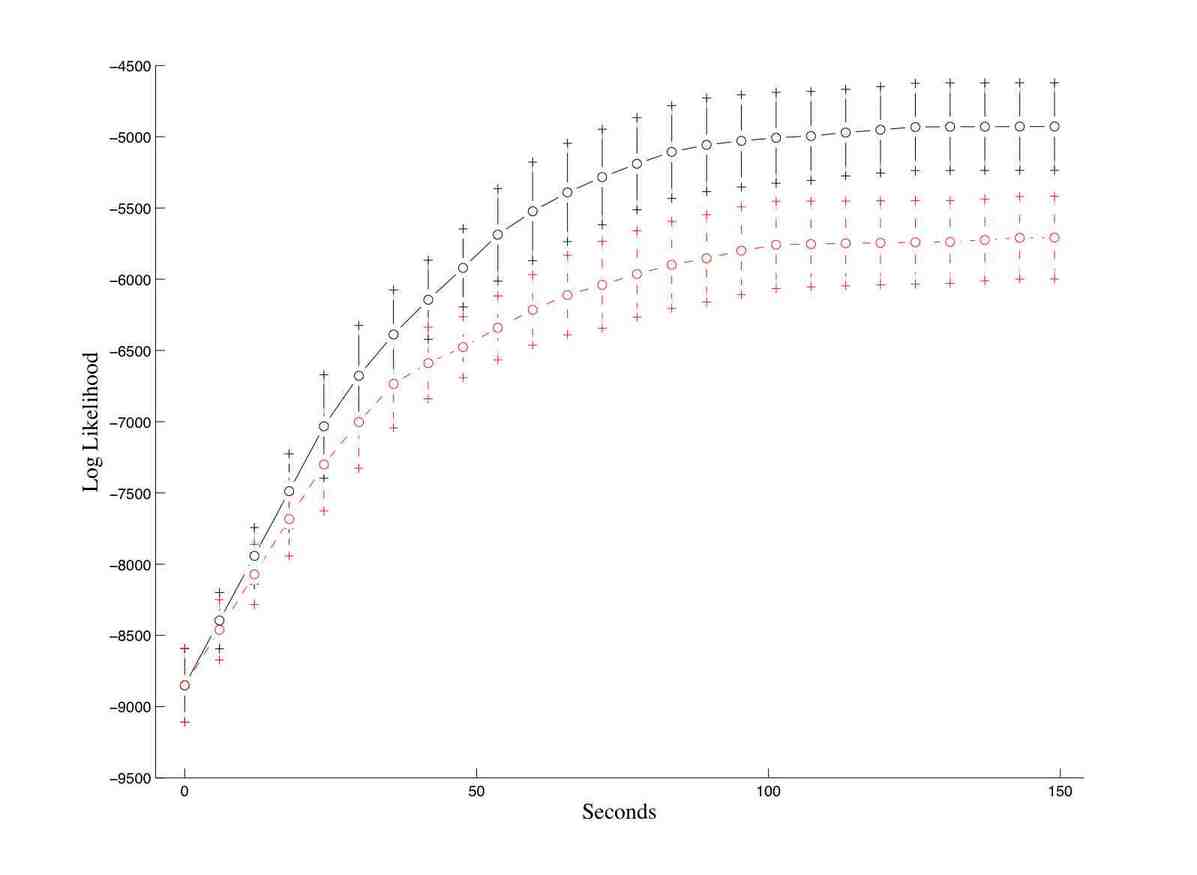}
 \hfill
 \includegraphics[width=7cm]{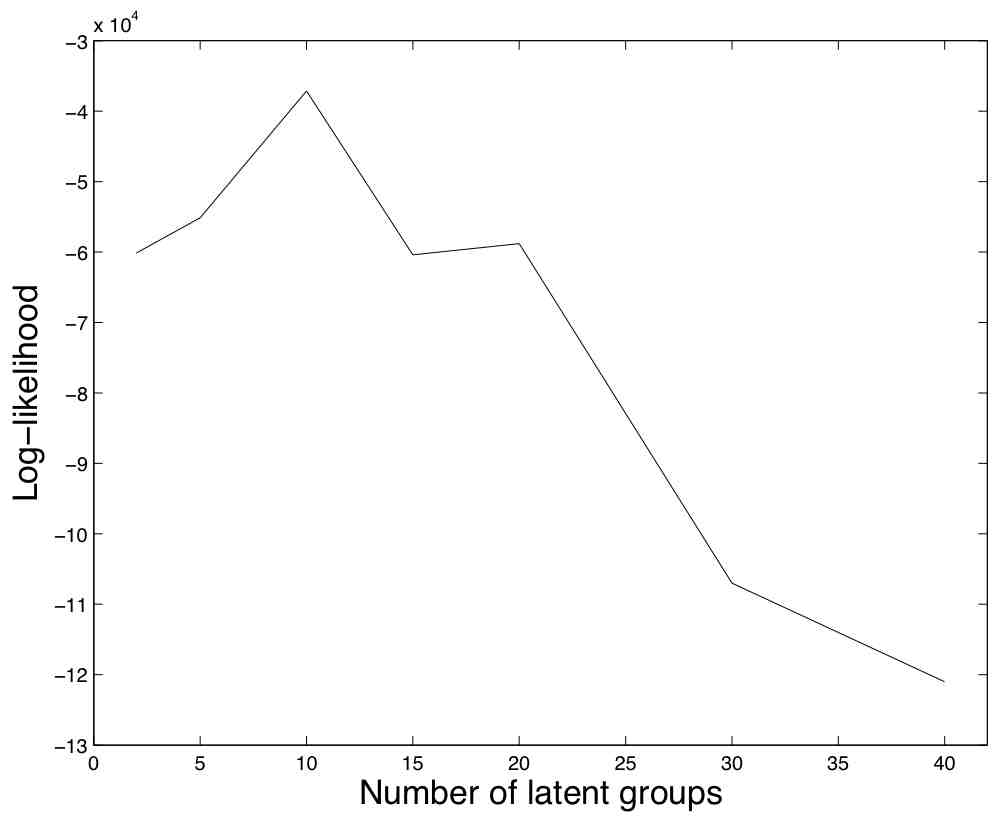}
 \caption{{\bf Left:}~The running time of the na\"ive variational inference (dashed, red line) against the running time of our enhanced (nested) variational inference algorithm (solid, black line), on a graph with 100 nodes and 4 clusters. {\bf Right:}~The held-out log-likelihood is indicative of the true number of latent clusters, on simulated data.  In the example shown, the peak identifies the correct number of clusters, $K^*=10$.} 
\label{fig:simdata1}
\end{figure}

 The variational EM algorithm successfully
recovers both the latent block model $B$ and the latent mixed
membership vectors $\vec{\pi}_{1:N}$.  In Figure \ref{fig:blk_mods} we
show the adjacency matrices of binary interactions where rows, i.e.,
nodes, are reordered according to their most likely membership. The nine panels are organized in to a three-by-three grid; panels in the same row correspond to the same combinations of (\# nodes and \# groups), whereas panels in the same columns correspond to the same value of $\alpha$ that was used to generate the data.  In each panel, the
estimated reordering of the nodes (i.e., the reordering of rows and columns in the interaction matrix) reveals the block model that was originally used
to simulate the interactions.  As $\alpha$ increases, each node is
likely to belong to more clusters.  As a consequence, they express
interaction patterns of clusters.  This phenomenon reflects in the
reordered interaction matrices as the block structure is less evident.
%

\subsubsection{Nested Variational Inference} 

 The nested variational
algorithm drives the log-likelihood to converge as fast as the na\"ive variational inference algorithm does, but reaches a significantly higher plateau.  In the left panel of Figure \ref{fig:simdata1}, we compare the running times of the nested variational-EM algorithm versus the
na\"ive implementation on a graph with 100 nodes and 4 clusters. We measure the number of seconds on the $X$ axis and the log-likelihood on the $Y$ axis. The two curves are averages over 26 experiments, and the error bars are at three standard deviations. Each of the 26 pairs of experiments was initialized with the same values for the parameters. The nested algorithm, which is more efficient in terms of space, converged faster.  Furthermore, the nested
variational algorithm can be parallelized given that the updates for
each interaction $(i,j)$ are independent of one another.

\subsubsection{Choosing the Number of Groups}

 Figure \ref{fig:simdata1} (right panel) shows an example where cross-validation is sufficient to perform model selection for the MMSB.  The example shown corresponds to a network among 300 nodes with $K=10$ clusters. We measure the number of latent clusters on the $X$ axis and the average held-out log-likelihood, corresponding to five-fold cross-validation experiments, on the $Y$ axis. A peak in this curve identifies the optimal number of clusters, to the extend of describing the data. 
 The nested variational EM algorithm was run till convergence, for each value of $K$ we tested, with a tolerance of $\epsilon=10^{-5}$. In the example shown, our estimate for $K$ occurs at the peak in the average held-out log-likelihood, and equals the correct number of clusters, $K^*=10$

\subsection{Application to Social Network Analysis}
\label{sec:adolescent_data}

 The National Longitudinal Study of Adolescent Health is nationally representative study that explores the how social contexts such as families, friends, peers, schools, neighborhoods, and communities influence health and risk behaviors of adolescents, and their outcomes in young adulthood \citep{Harr:Flor:Tabo:Bear:2003,Udry:2003}.
 Here, we analyze a friendship network among the students, at the same school that was considered by \cite{Hand:Raft:Tant:2006} and discussants.

\begin{figure}[ht!]
 \centering
  \includegraphics[width=0.90\textwidth]{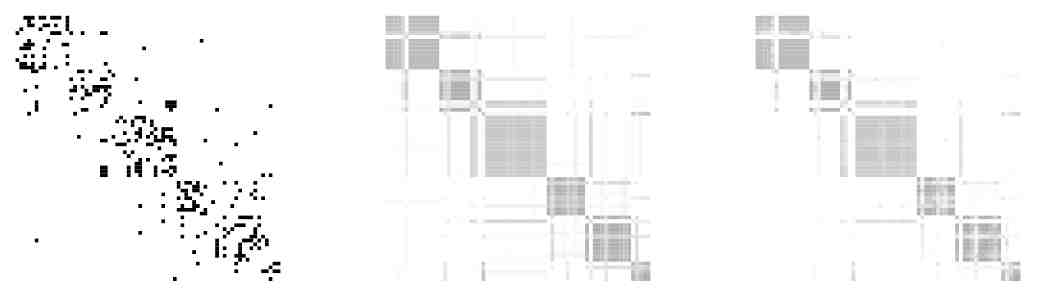}
  \caption{Original matrix of friensdhip relations among 69 students in grades 7 to 12 (left), and friendship estimated relations obtained by thresholding the posterior expectations $\vec{\pi}_p \,' B \,\vec{\pi}_q | R $ (center), and $\vec{\phi}_p \,' B \,\vec{\phi}_q | R$ (right).}
 \label{fig:add_health:data_exp}
\end{figure}

 A questionnaire was administered to a sample of students who were allowed to nominate up to 10 friends. At the school we picked, friendship nominations were collected among 71 students in grades 7 to 12. Two students did not nominate any friends so we analyzed the network of binary, asymmetric friendship relations among the remaining 69 students.
 The left panel of Figure \ref{fig:add_health:data_exp} shows the raw friendship relations, and we contrast this to the estimated networks in the central and right panels based on our model estimates.

\begin{figure}[t!]
 \centering
  \includegraphics[width=0.90\textwidth]{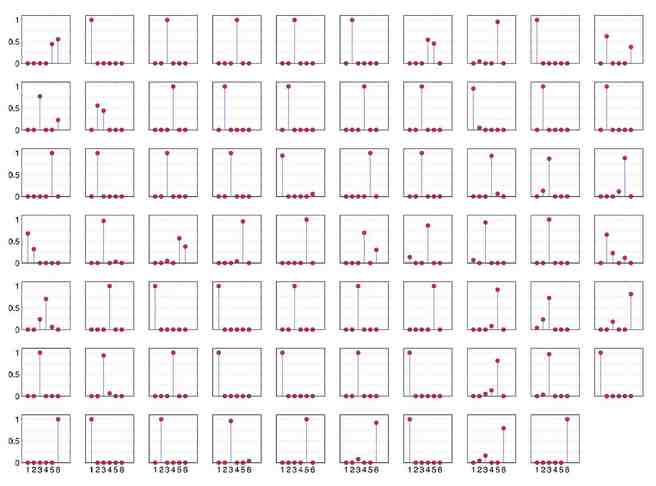}
  \caption{The posterior mixed membership scores, $\vec{\pi}$, for the 69 students.  Each panel correspond to a student; we order the clusters 1 to 6 on the $X$ axis, and we measure the student's grade of membership to these clusters on the $Y$ axis.}
 \label{fig:add_health:pi}
\end{figure}

 Given the size of the network we used BIC to perform model selection, as in the monks example of Section \ref{sec:monk_data}. The results suggest a model with $K^*=6$ groups. (We fix $K^*=6$ in the analyses that follow.)
 The hyper-parameters were estimated with the nested variational EM. They are $\hat{\alpha}=0.0487$, $\hat{\rho}=0.936$, and a fairly diagonal block-to-block connectivity matrix,
\[
 \hat B =
 \left[
  \bv{llllll}
    0.3235~ &  0.0     &  0.0     &  0.0     &  0.0     &  0.0    \\
    0.0     &  0.3614~ &  0.0002~ &  0.0     &  0.0     &  0.0    \\
    0.0     &  0.0     &  0.2607  &  0.0     &  0.0     &  0.0002 \\
    0.0     &  0.0     &  0.0     &  0.3751~ &  0.0009~ &  0.0    \\
    0.0     &  0.0     &  0.0     &  0.0002  &  0.3795  &  0.0    \\
    0.0     &  0.0     &  0.0     &  0.0     &  0.0     &  0.3719
  \ev
 \right].
\]
 Figure \ref{fig:add_health:pi} shows the expected posterior mixed membership scores for the 69 students in the sample; few students display mixed membership. The rarity of mixed membership in this context is expected. Mixed membership, instead, may signal unexpected social situations for further investigation. For instance, it may signal a family bond such as brotherhood, or a kid that is repeating a grade and is thus part of a broader social clique.
\begin{table}[t!]
\begin{center}
\begin{tabular}{r|rrrrrr|rrrrrr|rrrrrr}
       & \multicolumn{6}{c|}{MMSB Clusters} & \multicolumn{6}{c|}{MSB Clusters} & \multicolumn{6}{c}{LSCM Clusters} \\
 Grade &  1 & ~2 &  3 &  4 &  5 & ~6 &   1 & ~2 &  3 &  4 &  5 & ~6 &   1 & ~2 &  ~3 &  ~4 &  ~5 & ~6 \\ \hline
     7 & 13 &  1 &  0 &  0 &  0 &  0 &  13 &  1 &  0 &  0 &  0 &  0 &  13 &  1 &   0 &   0 &   0 &  0 \\
     8 &  0 &  9 &  2 &  0 &  0 &  1 &   0 & 10 &  2 &  0 &  0 &  0 &   0 & 11 &   1 &   0 &   0 &  0 \\
     9 &  0 &  0 & 16 &  0 &  0 &  0 &   0 &  0 & 10 &  0 &  0 &  6 &   0 &  0 &   7 &   6 &   3 &  0 \\
    10 &  0 &  0 &  0 & 10 &  0 &  0 &   0 &  0 &  0 & 10 &  0 &  0 &   0 &  0 &   0 &   0 &   3 &  7 \\
    11 &  0 &  0 &  1 &  0 & 11 &  1 &   0 &  0 &  1 &  0 & 11 &  1 &   0 &  0 &   0 &   0 &   3 & 10 \\
    12 &  0 &  0 &  0 &  0 &  0 &  4 &   0 &  0 &  0 &  0 &  0 &  4 &   0 &  0 &   0 &   0 &   0 &  4 \\
\end{tabular}
\end{center}
\caption{Grade levels versus (highest) expected posterior membership for the 69 students, according to three alternative models. MMSB is the proposed mixed membership stochastic blockmodel, MSB is a simpler stochastic block mixture model \citep{Dore:Bata:Ferl:2007}, and LSCM is the latent space cluster model \citep{Hand:Raft:Tant:2006}.}
\label{tab:add_health}
\end{table}
 In this data set we can successfully attempt an interpretation of the clusters in terms of grades. Table \ref{tab:add_health} shows the correspondence between clusters and grades in terms of students, for three alternative models. The three models are our mixed membership stochastic blockmodel (MMSB), a simpler stochastic block mixture model \citep{Dore:Bata:Ferl:2007} (MSB), and the latent space cluster model \citep{Hand:Raft:Tant:2006} (LSCM).

 Concluding this example, we note how the model decouples the observed friendship patterns into two complementary sources of variability. On the one hand, the connectivity matrix $B$ is a global, unconstrained set of hyper-parameters. On the other hand, the mixed membership vectors $\vec{\pi}_{1:N}$ provide a collection of node-specific latent vectors, which inform the directed connections in the graph in a symmetric fashion---and can be used to produce node-specific predictions.

\subsection{Application to Protein Interactions in \emph{Saccharomyces Cerevisiae}}
\label{sec:protein_data}

Protein-protein interactions (PPI) form the physical basis for the
formation of complexes and pathways that carry out different
biological processes.  A number of high-throughput experimental
approaches have been applied to determine the set of interacting
proteins on a proteome-wide scale in yeast. These include the
two-hybrid (Y2H) screens and mass spectrometry methods.  Mass spectrometry can be used to identify components of protein complexes
\citep{Gavi:Bosc:Krau:etal:2002,Ho:Gruh:Heil:etal:2002}.

High-throughput methods, though, may miss complexes that are not
present under the given conditions.  For example, tagging may disturb
complex formation and weakly associated components may dissociate and
escape detection.  Statistical models that encode information about
functional processes with high precision are an essential tool for
carrying out probabilistic de-noising of biological signals from
high-throughput experiments.

Our goal is to identify the proteins' diverse functional roles by
analyzing their local and global patterns of interaction via MMSB.  The
biochemical composition of individual proteins make them suitable for
carrying out a specific set of cellular operations, or
\textit{functions}.  Proteins typically carry out these functions as
part of stable protein complexes \citep{Krog:Cagn:Yu:Zhon:2006}.
There are many situations in which proteins are believed to interact
\citep{Albe:John:Lewi:Raff:2002}. The main intuition behind our
methodology is that pairs of protein interact because they are part of
the same stable protein complex, i.e., co-location, or because they
are part of interacting protein complexes as they carry out compatible
cellular operations.

\subsubsection{Gold Standards for Functional Annotations}

\begin{table}[b!]
\begin{center}
\begin{tabular}{rlr|rlr}
\# & Category                           & Count & \# & Category                           & Count \\ \hline
 1 & Metabolism                         &   125 &  9 & Interaction w/ cell. environment   &    18 \\
 2 & Energy                             &    56 & 10 & Cellular regulation                &    37 \\
 3 & Cell cycle \& DNA processing       &   162 & 11 & Cellular other                     &    78 \\
 4 & Transcription (tRNA)               &   258 & 12 & Control of cell organization       &    36 \\
 5 & Protein synthesis                  &   220 & 13 & Sub-cellular activities            &   789 \\
 6 & Protein fate                       &   170 & 14 & Protein regulators                 &     1 \\
 7 & Cellular transportation            &   122 & 15 & Transport facilitation             &    41 \\
 8 & Cell rescue, defence \& virulence  &     6 &    &                                    &         \\
\end{tabular}
\end{center}
\caption{The 15 high-level functional categories obtained by cutting the MIPS annotation tree at the first level and how many proteins (among the 871 we consider) participate in each of them. Most proteins participate in more than one functional category, with an average of $\approx 2.4$ functional annotations for each protein.}
\label{tab:functions}
\end{table}

 The Munich Institute for Protein Sequencing (MIPS) database was
created in 1998 based on evidence derived from a variety of
experimental techniques, but does not include information from
high-throughput data sets \citep{Mewe:Amid:Arno:etal:2004}.
 It contains about 8000 protein complex associations in yeast.  We analyze
a subset of this collection containing 871 proteins, the interactions
amongst which were hand-curated. The institute also provides a set of
functional annotations, alternative to the gene ontology (GO).  These
annotations are organized in a tree, with 15 general functions at the
first level, 72 more specific functions at an intermediate level, and
255 annotations at the the leaf level.  In Table \ref{tab:functions}
we map the 871 proteins in our collections to the main functions of
the MIPS annotation tree; proteins in our sub-collection have about
$2.4$ functional annotations on average.\footnote{We note that the
  relative importance of functional categories in our sub-collection,
  in terms of the number of proteins involved, is different from the
  relative importance of functional categories over the entire MIPS
  collection.}

\begin{figure}[t!]
  \centering
   \includegraphics[width=15cm]{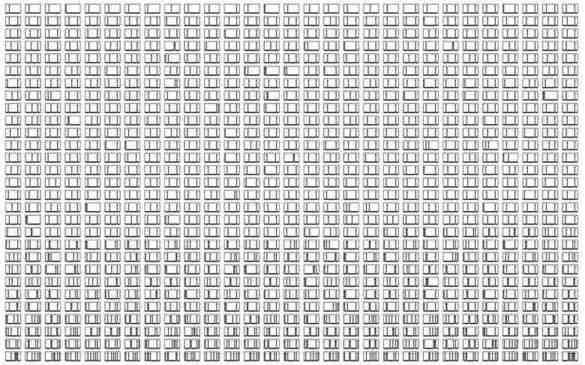}
  \centering
   \includegraphics[width=15cm]{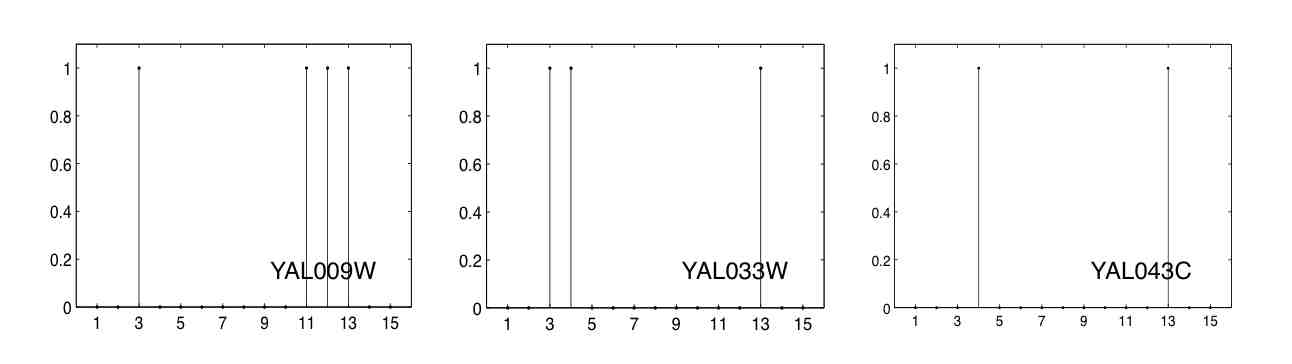}
  \caption{By mapping individual proteins to the 15 general
    functions in Table \ref{tab:functions}, we obtain a 15-dimensional representation for each
    protein.  Here, each panel corresponds to a protein; the
    15 functional categories are displayed on the $X$ axis, whereas
    the presence or absence of the corresponding functional annotation
    is displayed on the $Y$ axis.  The plots at the bottom zoom into three example panels (proteins).}
\label{fig:mipsdata}
\end{figure}

By mapping proteins to the 15 general functions, we obtain a
15-dimensional representation for each protein.  In Figure
\ref{fig:mipsdata} each panel corresponds to a protein; the 15
functional categories are ordered as in Table \ref{tab:functions} on
the $X$ axis, whereas the presence or absence of the corresponding
functional annotation is displayed on the $Y$ axis.

\subsubsection{Brief Summary of Previous Findings}
\label{sec:previously}

In previous work, we established the usefulness of an admixture of
latent blockmodels for analyzing protein-protein interaction data
\citep{Airo:Blei:Xing:Fien:2005}. For example, we used the MMSB for
testing functional interaction hypotheses (by setting a null
hypothesis for $B$), and unsupervised estimation experiments. In the next Section, we assess whether, and how much, functionally relevant biological signal can be captured in by the MMSB.

In summary, the results in \citet{Airo:Blei:Xing:Fien:2005} show that the MMSB identifies protein complexes
whose member proteins are tightly interacting with one another.  The
identifiable protein complexes correlate with the following four
categories of Table \ref{tab:functions}: cell cycle \& DNA processing,
transcription, protein synthesis, and sub-cellular activities.  The
high correlation of inferred protein complexes can be leveraged for
predicting the presence of absence of functional annotations, for
example, by using a logistic regression.  However, there is not enough
signal in the data to independently predict annotations in other
functional categories.  The empirical Bayes estimates of the
hyper-parameters that support these conclusions in the various types
of analyses are consistent; $\hat \alpha<1$ and small; and $\hat B$
nearly block diagonal with two positive blocks comprising the four
identifiable protein complexes.  In these previous analyses, we fixed the
number of latent protein complexes to 15; the number of broad functional categories in Table \ref{tab:functions}.

\begin{figure}[b!]
  \centering
   \includegraphics[width=7.5cm]{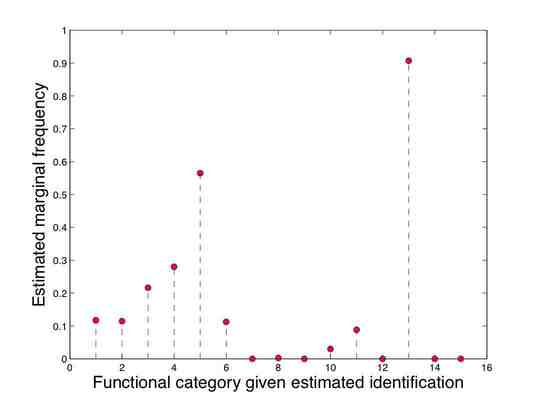}
   \includegraphics[width=7.5cm]{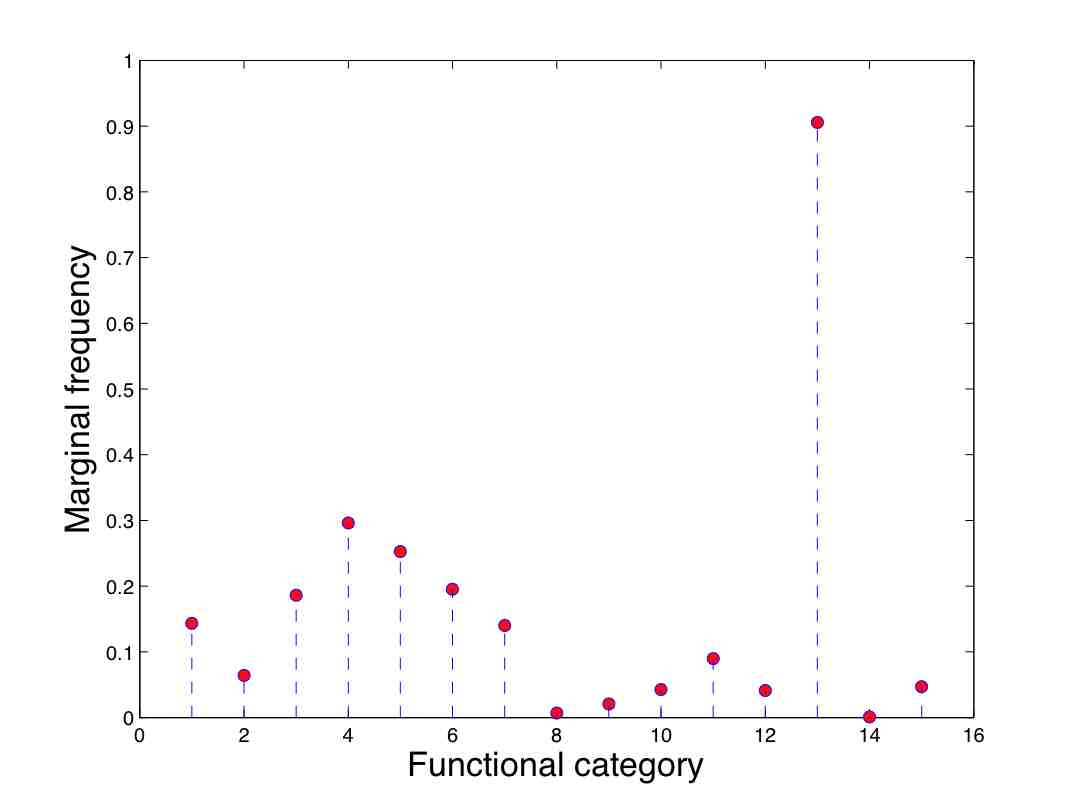}
  \caption{We estimate the mapping of latent groups to functions.  The
    two plots show the marginal frequencies of membership of proteins
    to true functions (bottom) and to identified functions (top), in
    the cross-validation experiment.  The mapping is selected to
    maximize the accuracy of the predictions on the training set, in
    the cross-validation experiment, and to minimize the divergence
    between marginal true and predicted frequencies if no training
    data is available---see Section \ref{sec:previously}.}
\label{fig:identify}
\end{figure}
The latent protein complexes are not a-priori identifiable in our
model.  To resolve this, we estimated a mapping between latent complexes and
functions by minimizing the divergence between true and predicted
marginal frequencies of membership, where the truth was evaluated on a
small fraction of the interactions.  We used this mapping to compare
predicted versus known functional annotations for all proteins.  The
best estimated mapping is shown in the left panel of Figure \ref{fig:identify}, along with the marginal latent category membership, and it is compared to the 15 broad functional categories Table \ref{tab:functions}, along with the known category membership (in the MIPS database), in the right panel.
\begin{figure}[t!]
  \centering
   \includegraphics[width=14cm]{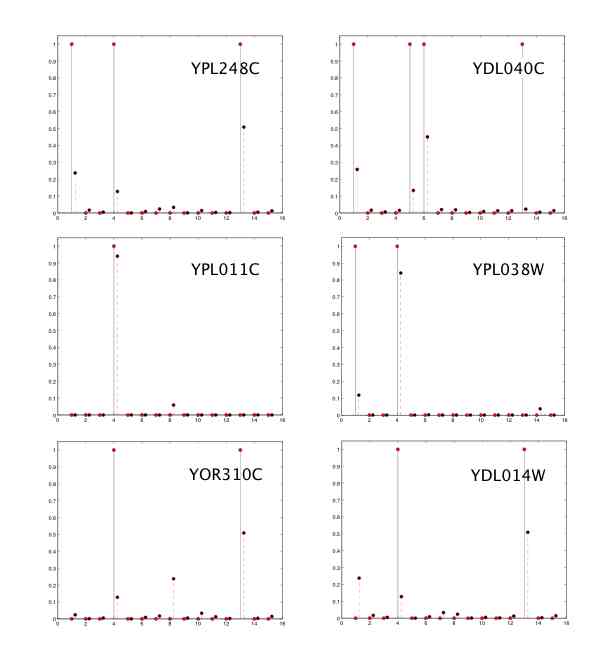}
 \caption{Predicted mixed-membership probabilities (dashed, red lines)
   versus binary manually curated functional annotations (solid, black
   lines) for 6 example proteins.  The identification of latent groups
   to functions is estimated, Figure
   \ref{fig:identify}.}
\label{fig:exampleprot}
\end{figure}
 Figure \ref{fig:exampleprot} displays a few examples of predicted
mixed membership probabilities against the true annotations, given the
\textit{estimated mapping} of latent protein complexes to functional
categories.

\subsubsection{Measuring the Functional Content in the Posterior}
\label{sec:functional_content}

 In a follow-up study we considered the gene ontology (GO) \citep{Ashb:etal:2000} as the source of functional annotations to consider as ground truth in our analyses. GO is a broader and finer grained functional annotation scheme if compared to that produced by the Munich Institute for Protein Sequencing.
 Furthermore, we explored a much larger model space than in the previous study, in order to tests to what extent MMSB can reduce the dimensionality of the data while revealing substantive information about the functionality of proteins that can be used to inform subsequent analyses.
 We fit models with a number blocks up to $K=225$. Thanks to our nested variational inference algorithm, we were able to perform five-fold cross-validation for each value of $K$. We determined that a fairly parsimonious model $(K^*=50)$ provides a good description of the observed protein interaction network. This fact is (qualitatively) consistent with the quality of the predictions that were obtained with a parsimonious model ($K=15$) in the previous section, in a different setting. This finding supports the hypothesis that groups of interacting proteins in the MIPS data set encode biological signal at a scale of aggregation that is higher than that of protein complexes.\footnote{It has been recently suggested that stable protein complexes average five proteins in size \citep{Krog:Cagn:Yu:Zhon:2006}. Thus, if MMSB captured biological signal at the protein-complex resolution, we would expect the optimal number of groups to be much higher (Disregarding mixed membership, $871/5\approx 175$.)}
\begin{figure}[b!]
  \centering
   \includegraphics[width=16.5cm]{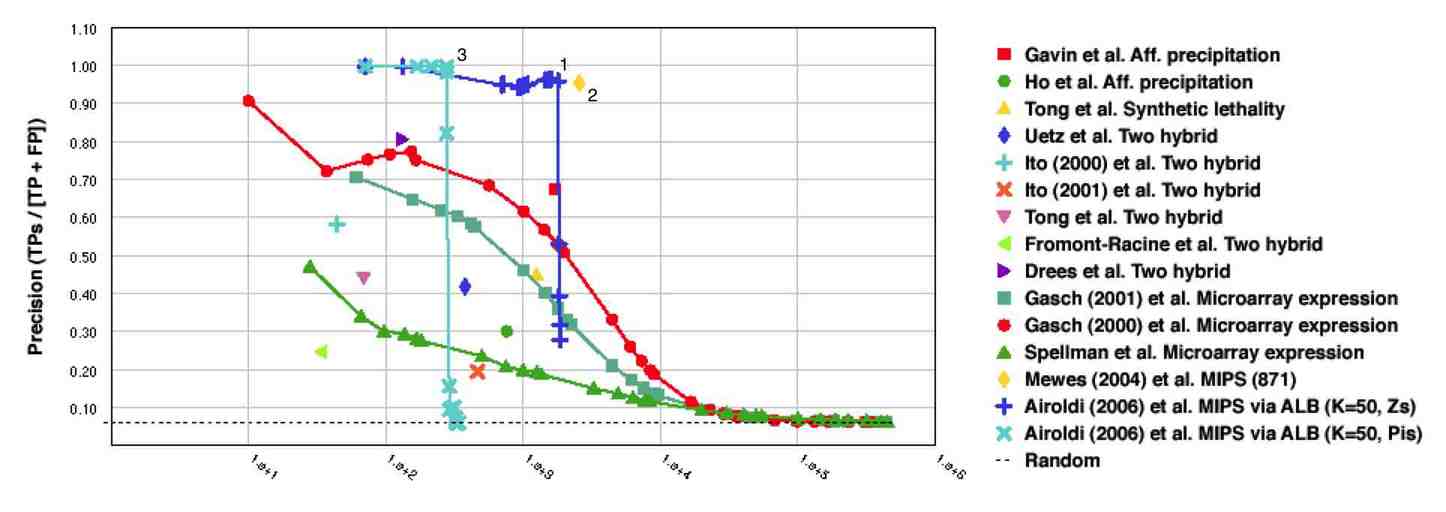}
  \caption{In the top panel we measure the functional content of the
    the MIPS collection of protein interactions (yellow diamond), and
    compare it against other published collections of interactions and
    microarray data, and to the posterior estimates of the MMSB
    models---computed as described in Section
    \ref{sec:functional_content}. A breakdown of three estimated
    interaction networks (the points annotated 1, 2, and 3) into most represented
    gene ontology categories is detailed in Table
    \ref{tab:detail_functional_content}.}
\label{fig:functional_content}
\end{figure}

 We settled on a model with $K^*=50$ blocks.
 To evaluate the functional content of the interactions predicted by such model, we first computed the posterior probabilities of interactions by thresholding the posterior expectations
\[
\mathbb{E} \bigm[ R(p,q)=1 \bigm] \approx \widehat{\vec{\pi}}_p\,' \,
\widehat{B} ~ \, \widehat{\vec{\pi}}_q \qquad \hbox{ and } \qquad
\mathbb{E} \bigm[ R(p,q)=1 \bigm] \approx \widehat{\vec{\phi}}_{\ptoq}\,' \,
\widehat{B} ~ \, \widehat{\vec{\phi}}_{\pfromq},
\]
 and we then computed the precision-recall curves corresponding to these predictions \citep{Myer:Barr:Hibb:Hutt:2006}. These curves are shown in Figure \ref{fig:functional_content} as the light blue ($-\times$) line and the the dark blue ($-+$) line.
 In Figure \ref{fig:functional_content} we also plotted the functional content of the original MIPS collection.
\begin{table}[b!]
\begin{center}
\begin{tabular}{rlp{9cm}rr} \hline
\# & GO Term    & Description                                   & Pred. & Tot.   \\ \hline
 1 & GO:0043285 & Biopolymer catabolism                         &   561 &  17020 \\
 1 & GO:0006366 & Transcription from RNA polymerase II promoter &   341 &  36046 \\
 1 & GO:0006412 & Protein biosynthesis                          &   281 & 299925 \\
 1 & GO:0006260 & DNA replication                               &   196 &   5253 \\
 1 & GO:0006461 & Protein complex assembly                      &   191 &  11175 \\
 1 & GO:0016568 & Chromatin modification                        &   172 &  15400 \\
 1 & GO:0006473 & Protein amino acid acetylation                &    91 &    666 \\
 1 & GO:0006360 & Transcription from RNA polymerase I promoter  &    78 &    378 \\
 1 & GO:0042592 & Homeostasis                                   &    78 &   5778 \\ \hline
 2 & GO:0043285 & Biopolymer catabolism                         &   631 &  17020 \\
 2 & GO:0006366 & Transcription from RNA polymerase II promoter &   414 &  36046 \\
 2 & GO:0016568 & Chromatin modification                        &   229 &  15400 \\
 2 & GO:0006260 & DNA replication                               &   226 &   5253 \\
 2 & GO:0006412 & Protein biosynthesis                          &   225 & 299925 \\
 2 & GO:0045045 & Secretory pathway                             &   151 &  18915 \\
 2 & GO:0006793 & Phosphorus metabolism                         &   134 &  17391 \\
 2 & GO:0048193 & Golgi vesicle transport                       &   128 &   9180 \\
 2 & GO:0006352 & Transcription initiation                      &   121 &   1540 \\ \hline
 3 & GO:0006412 & Protein biosynthesis                          &   277 & 299925 \\
 3 & GO:0006461 & Protein complex assembly                      &   190 &  11175 \\
 3 & GO:0009889 & Regulation of biosynthesis                    &    28 &    990 \\
 3 & GO:0051246 & Regulation of protein metabolism              &    28 &    903 \\
 3 & GO:0007046 & Ribosome biogenesis                           &    10 &  21528 \\
 3 & GO:0006512 & Ubiquitin cycle                               &     3 &   2211 \\ \hline
\end{tabular}
\end{center}
\caption{Breakdown of three example interaction networks into most represented gene ontology categories---see text for more details. The digit in the first column indicates the example network in Figure \ref{fig:functional_content} that any given line refers to. The last two columns quote the number of predicted, and possible pairs for each GO term.}
\label{tab:detail_functional_content}
\end{table}
 This plot confirms that the MIPS collection of interactions, our data, is one of the most precise (the $Y$ axis measures precision) and most extensive (the $X$ axis measures the amount of functional annotations predicted, a measure of recall) source of biologically relevant interactions available to date---the yellow diamond, point \# 2.
\begin{sidewaysfigure}
  \centering
   \includegraphics[width=21.5cm]{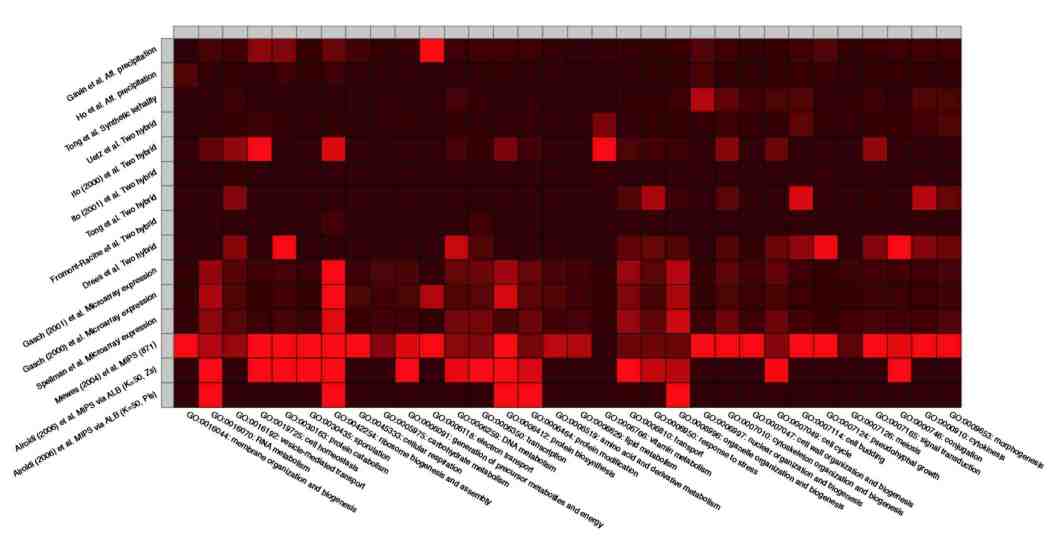}
  \caption[]{We investigate the correlations between data collections
    (rows) and a sample of gene ontology categories (columns). The
    intensity of the square (red is high) measures the area under the
    precision-recall curve.}
\label{fig:GO_correlations}
\end{sidewaysfigure}
 The posterior means of $(\vec{\pi}_{1:N})$ and the estimates of $(\alpha,B)$ provide a parsimonious representation for the MIPS collection, and lead to precise interaction estimates, in moderate amount (the light blue, $-\times$ line).  The posterior means of $(Z_\rightarrow,Z_\leftarrow)$ provide a richer representation for the data, and describe most of the functional content of the MIPS collection with high precision (the dark blue, $-+$ line).
 Most importantly, notice the estimated protein interaction networks, i.e., pluses and crosses, corresponding to lower levels of recall feature a more precise functional content than the original. This means that the proposed latent block structure is helpful in summarizing the collection of interactions---by ranking them properly. (It also happens that dense blocks of predicted interactions contain known functional predictions that were not in the MIPS collection.)
 Table \ref{tab:detail_functional_content} provides more information about three instances of predicted interaction networks displayed in Figure \ref{fig:functional_content}; namely, those corresponding the points annotated with the numbers 1 (a collection of interactions predicted with the $\vec{\pi}$'s), 2 (the original MIPS collection of interactions), and 3 (a collection of interactions predicted with the $\vec{\phi}$'s).
 Specifically, the table shows a breakdown of the predicted (posterior) collections of interactions in each example network into the gene ontology categories. A count in the second-to-last column of Table \ref{tab:detail_functional_content} corresponds to the fact that both proteins are annotated with the same GO functional category.\footnote{Note that, in GO, proteins are typically annotated to multiple functional categories.} 
 Figure \ref{fig:GO_correlations} investigates the correlations between the data sets (in rows) we considered in Figure \ref{fig:functional_content} and few gene ontology categories (in columns).  The intensity of the square (red is high) measures the area under the precision-recall curve \citep{Myer:Barr:Hibb:Hutt:2006}.

 In this application, the MMSB learned information about (i) the mixed membership of objects to latent groups, and (ii) the connectivity patterns among latent groups.  These quantities were useful in describing and summarizing the functional content of the MIPS collection of protein interactions. This suggests the use of MMSB as a dimensionality reduction approach that may be useful for performing model-driven de-noising of new collections of interactions, such as those measured via high-throughput experiments.

\section{Discussion}

 Below we place our research in a larger modeling context, offer some insights into the inner workings of the model, and briefly comment on limitations and extensions.


 Modern probabilistic models for relational data analysis are rooted in the stochastic blockmodels for psychometric and sociological analysis, pioneered by \citet{Lorr:Whit:1971} and by \citet{Holl:Lein:1975}.
 In statistics, this line of research has been extended in various contexts over the years \citep{Fien:Meye:Wass:1985,Wass:Patt:1996,Snij:2002,Hoff:Raft:Hand:2002,Dore:Bata:Ferl:2004}.
 In machine learning, the related technique of Markov random networks \citep{Fran:Stra:1986} have been used for link prediction \citep{Task:Wong:Abbe:Koll:2003} and the traditional blockmodels have been extended to include nonparametric Bayesian priors \citep{Kemp:Grif:Tene:2004,Kemp:Tene:Grif:etal:2006} and to integrate relations and text \citep{McCa:Wang:Moha:2007}.

  There is a particularly close relationship between the MMSB and the latent space models \citep{Hoff:Raft:Hand:2002,Hand:Raft:Tant:2006}.  In the latent space models, the latent vectors are drawn from Gaussian distributions and the interaction data is drawn from a Gaussian with mean $\vec{\pi}_p \,' \mathbb{I} \vec{\pi}_q$.  In the MMSB, the marginal probability of an interaction takes a similar form, $\vec{\pi}_p \,' B \vec{\pi}_q$, where $B$ is the matrix of probabilities of interactions for each pair of latent groups.
 Two major differences exist between these approaches. In MMSB, the distribution over the latent vectors is a Dirichlet and the underlying data distribution is arbitrary---we have chosen Bernoulli. The posterior inference in latent space models \citep{Hoff:Raft:Hand:2002,Hand:Raft:Tant:2006} is carried out via MCMC sampling, while we have developed a scalable variational inference algorithm to analyze large network structures. (It would be interesting to develop a variational algorithm for the latent space models as well.)


 We note how the model decouples the observed friendship patterns into two complementary sources of variability. On the one hand, the connectivity matrix $B$ is a global, unconstrained set of hyper-parameters. On the other hand, the mixed membership vectors $\vec{\pi}_{1:N}$ provide a collection of node-specific latent vectors, which inform the directed connections in the graph in a symmetric fashion. Last, the single membership indicators $(\vec{z}_{\ptoq}, \vec{z}_{\pfromq})$ provide a collection interaction-specific latent variables.

 A recurring question, which bears relevance to mixed membership models in general, is why we do not integrate out the single membership indicators---$(\vec{z}_{\ptoq}, \vec{z}_{\pfromq})$. While this may lead to computational efficiencies we would often lose interpretable quantities that are useful for making predictions, for de-noising new measurements, or for performing other tasks. In fact, the posterior distributions of such quantities typically carry substantive information about elements of the application at hand. In the application to protein interaction networks of Section \ref{sec:protein_data}, for example, they encode the interaction-specific memberships of individual proteins to protein complexes.


 A limitation of our model can be best appreciated in a simulation setting.
 If we consider structural properties of the network MMSB is capable of generating, we count a wide array of local and global connectivity patterns. But the model does not readily generate \textit{hubs}, that is, nodes connected with a large number of directed or undirected connections, or networks with skewed degree distributions.

 
 From a data analysis perspective, we speculate that the value of MMSB in capturing substantive information about a problem will increase in semi-supervised setting---where, for example, information about the membership of genes to functional contexts is included in the form of prior distributions. In such a setting we may be interested in looking at the change between prior and posterior membership; a sharp change may signal biological phenomena worth investigating.

 We need not assume that the number of groups/blocks, $K$, is finite. It is possible, for example, to posit that the mixed-membership vectors are sampled form a stochastic process $D_\alpha$, in the nonparametric setting. In particular, in order to maintain mixed membership of nodes to groups/blocks we need to sample them from a hierarchical Dirichlet process \citep{Teh:Jord:Beal:Blei:2006}, rather than from a Diriclet Process \citep{Esco:West:1995}.

\section{Conclusions}

In this paper we introduced mixed membership stochastic blockmodels, a
novel class of latent variable models for relational data.  These
models provide exploratory tools for scientific analyses in
applications where the observations can be represented as a collection
of unipartite graphs.  The nested variational inference algorithm is
parallelizable and allows fast approximate inference on large graphs.


 The relational nature of such data as well as the multiple goals of the analysis in the applications we considered motivated our technical choices. Latent variables in our models are introduced to capture application-specific substantive elements of interest, e.g., monks and factions in the monastery.
 The applications to social and biological networks we considered share considerable similarities in the way such elements relate.  This allowed us to identify a general formulation of the model that we present in Appendix \ref{app:general_formulation}.
 Approximate variational inference for the general model is presented in Appendix \ref{app:general_inference}. 

\vskip 0.2in

\subsection*{Acknowledgments}

 This work was partially supported by National Institutes of
  Health under Grant No. R01 AG023141-01, by the Office of Naval
  Research under Contract No. N00014-02-1-0973, by the National
  Science Foundation under Grants No. DMS-0240019, IIS-0218466, and
  DBI-0546594, by the Pennsylvania Department of Health's Health
  Research Program under Grant No. 2001NF-Cancer Health Research Grant
  ME-01-739, and by the Department of Defense, all to Carnegie Mellon
  University.

\bibliographystyle{plainnat}

\appendix
\section{General Model Formulation}
\label{app:general_formulation}

 In general, mixed membership stochastic blockmodels can be specified in terms of assumptions at four levels: population, node, latent variable, and sampling scheme level.

\subsubsection*{A1--Population Level}
 Assume that there are $K$ classes or sub-populations in the population of interest.
 We denote by $f\bigm(R(p,q) \mid B(g,h) \bigm)$ the probability distribution of the relation measured on the pair of nodes $(p,q)$, where the $p$-$th$ node is in the $h$-$th$ sub-population, the $q$-$th$ node is in the $h$-$th$ sub-population, and $B(g,h)$ contains the relevant parameters.  The indices $i,j$ run in $1,\dots,N$, and the indices $g,h$ run in $1,\dots,K$.

\subsubsection*{A2---Node Level}
 The components of the membership vector $\vec{\pi}_p = [\vec{\pi}_p(1), \dots, \vec{\pi}_p(k)]'$ encodes the mixed membership of the $n$-$th$ node to the various sub-populations.
 The distribution of the observed response $R(p,q)$ given the relevant, node-specific memberships, $(\vec{\pi}_p, \vec{\pi}_q)$, is then
\begin{equation}
 \label{eq:mixture}
 Pr \bigm( R(p,q) \mid \vec{\pi}_p, \vec{\pi}_q, B \bigm) ~=~ \sum_{g,h=1}^K \vec{\pi}_p(g) ~ f(R(p,q) \mid B(g,h)) ~ \vec{\pi}_q(h).
\end{equation}
 Conditional on the mixed memberships, the response edges $y_{jnm}$  are independent of one another, both across distinct graphs and pairs of nodes.

\subsubsection*{A3---Latent Variable Level}
 Assume that the mixed membership vectors $\vec{\pi}_{1:N}$ are realizations of a latent variable with distribution $D_{\vec{\alpha}}$, with parameter vector $\vec{\alpha}$.  The probability of observing $R(p,q)$, given the parameters, is then
\begin{equation}
 Pr \bigm( R(p,q) \mid \vec{\alpha}, B \bigm) ~=~ \int   Pr \bigm( R(p,q) \mid \vec{\pi}_p, \vec{\pi}_q, B \bigm)  ~ D_{\vec\alpha}(d\vec\pi).
\end{equation}

\subsubsection*{A4---Sampling Scheme Level}
 Assume that the $M$ independent replications of the relations measured on the population of nodes are independent of one another.  The probability of observing the whole collection of graphs, $R_{1:M}$, given the parameters, is then given by the following equation.
 \begin{equation}
  \label{eq:latent}
 Pr \bigm( R_{1:M} \mid \vec{\alpha}, B \bigm) \, =  \prod_{m=1}^M \prod_{p,q=1}^N Pr \bigm( R_m(p,q) \mid \vec{\alpha}, B \bigm).
 \end{equation}

 Full model specifications immediately adapt to the different kinds of data, e.g., multiple data types through the choice of $f$, or parametric or semi-parametric specifications of the prior on the number of clusters through the choice of $D_\alpha$.

\section{Details of the Variational Approximation}
\label{app:general_inference}

 Here we present more details about the derivation of the variational EM algorithm presented in Section \ref{sec:estimation_inference}. Furthermore, we address a setting where $M$ replicates are available about the paired measurements, $G_{1:M}=(N,R_{1:M})$, and relations $R_m(p,q)$ take values into an arbitrary metric space according to $f \bigm( R_m(p,q) \mid .. \bigm)$.
 An extension of the inference algorithm to address the case or multivariate relations, say $J$-dimensional, and multiple blockmodels $B_{1:J}$ each corresponding to a distinct relational response, can be derived with minor modifications of the derivations that follow.

\subsection{Variational Expectation-Maximization}

 We begin by briefly summarizing the general strategy we intend to use.
 The approximate variant of EM we describe here is often referred to as {\it Variational EM} \citep{Beal:Ghah:2003}. We begin by rewriting  $Y=R$ for the data, $X=(\vec{\pi}_{1:N},Z_\rightarrow,Z_\leftarrow)$ for the latent variables, and $\Theta=(\vec{\alpha},B)$ for the model's parameters.
 Briefly, it is possible to lower bound the likelihood, $p(Y|\Theta)$, making use of Jensen's inequality and of any distribution on the latent variables $q(X)$,
\begin{eqnarray}
 p(Y|\Theta) & =    & \log \int_\mathcal{X} ~ p(Y,X|\Theta) ~ dX \nonumber \\
             & =    & \log \int_\mathcal{X} ~ q(X) ~ \frac{p(Y,X|\Theta)}{q(X)} ~ dX \qquad \hbox{(for any $q$)} \nonumber \\
             & \geq & \int_\mathcal{X} ~ q(X) ~ \log \frac{p(Y,X|\Theta)}{q(X)} ~ dX \qquad \hbox{(Jensen's)} \nonumber \\
             & =    & \mathbb{E}_q \bigm[ \log p(Y,X|\Theta) - \log q(X) \bigm] ~ =: ~ \mathcal{L}(q,\Theta)
\end{eqnarray}
 In EM, the lower bound $\mathcal{L}(q,\Theta)$ is then iteratively maximized with respect to $\Theta$, in the M step, and $q$ in the E step \citep{Demp:Lair:Rubi:1977}.  In particular, at the $t$-$th$ iteration of the E step we set
\begin{equation}
\label{eq:em_post}
 q^{(t)} = p(X|Y,\Theta^{(t-1)}),
\end{equation}
 that is, equal to the posterior distribution of the latent variables given the data and the estimates of the  parameters at the previous iteration.

 Unfortunately, we cannot compute the posterior in Equation \ref{eq:em_post} for the admixture of latent blocks model.  Rather, we define a direct parametric approximation to it, $\tilde q = q_\Delta (X)$, which involves an extra set of {\it variational parameters}, $\Delta$, and entails an approximate lower bound for the likelihood $\mathcal{L}_\Delta (q,\Theta)$.
 At the $t$-$th$ iteration of the E step, we then minimize the Kullback-Leibler divergence between $q^{(t)}$ and $q^{(t)}_\Delta$, with respect to $\Delta$, using the data.\footnote{This is equivalent to maximizing the approximate lower bound for the likelihood, $\mathcal{L}_\Delta (q,\Theta)$, with respect to $\Delta$.}  The optimal parametric approximation is, in fact, a proper posterior as it depends on the data $Y$, although indirectly, $q^{(t)} \approx q_{\Delta^*(Y)}^{(t)} (X) = p (X|Y) $.

\subsection{Lower Bound for the Likelihood}

According to the mean-field theory \citep{Jord:Ghah:Jaak:Saul:1999,Xing:Jord:Russ:2003}, one can approximate an intractable distribution such as the one defined by Equation (\ref{eq:likelihood_with_zs}) by a fully factored distribution $q(\vec{\pi}_{1:N},Z^\rightarrow_{1:M},Z^\leftarrow_{1:M})$ defined as follows:
\begin{eqnarray}
  & & q(\vec{\pi}_{1:N},Z^\rightarrow_{1:M},Z^\leftarrow_{1:M} | \vec{\gamma}_{1:N},\Phi^\rightarrow_{1:M},\Phi^\leftarrow_{1:M}) \nonumber \\
  & = & \prod_{p} ~ q_1 (\vec{\pi}_p | \vec{\gamma}_p) ~ \prod_{m} \prod_{p,q} ~ \Big( q_2 (\vec{z}_{\ptoq}^m | \vec{\phi}_{\ptoq}^m,1) ~ q_2 (\vec{z}_{\pfromq}^m | \vec{\phi}_{\pfromq}^m,1) \Big), \qquad
\end{eqnarray}
 where $q_1$ is a Dirichlet, $q_2$ is a multinomial, and $\Delta = (\vec{\gamma}_{1:N},\Phi^\rightarrow_{1:M},\Phi^\leftarrow_{1:M})$ represent the set of free {\em variational parameters} need to be estimated in the approximate distribution.

Minimizing the Kulback-Leibler divergence between this $q( \vec{\pi}_{1:N},Z^\rightarrow_{1:M},Z^\leftarrow_{1:M} |\Delta)$ and the original $p( \vec{\pi}_{1:N},Z^\rightarrow_{1:M},Z^\leftarrow_{1:M}$ defined by Equation (\ref{eq:likelihood_with_zs}) leads to the following approximate lower bound for the likelihood.
{\small
\bvq
 \mathcal{L}_\Delta (q,\Theta)
 & = & \mathbb{E}_q \bigm[ \log \prod_m \prod_{p,q} ~ p_1 ( R_m(p,q) | \vec{z}_{\ptoq}^m, \vec{z}_{\pfromq}^m, B) \bigm]  \nonumber \\
 & + & \mathbb{E}_q \bigm[ \log \prod_m \prod_{p,q} ~ p_2 (\vec{z}_{\ptoq}^m | \vec{\pi}_p,1) \bigm] + \mathbb{E}_q \bigm[ \log \prod_m \prod_{p,q} ~ p_2 (\vec{z}_{\pfromq}^m | \vec{\pi}_q,1) \bigm]  \nonumber \\
 & + & \mathbb{E}_q \bigm[ \log \prod_p ~ p_3 (\vec{\pi}_p | \vec{\alpha}) \bigm] - \mathbb{E}_q \bigm[ \prod_{p} ~ q_1 (\vec{\pi}_p | \vec{\gamma}_p) \bigm]  \nonumber \\
 & - & \mathbb{E}_q \bigm[ \log \prod_{m} \prod_{p,q} ~ q_2 (\vec{z}_{\ptoq}^m | \vec{\phi}_{\ptoq}^m,1)\bigm] - \mathbb{E}_q \bigm[ \log \prod_{m} \prod_{p,q} ~ q_2 (\vec{z}_{\pfromq}^m | \vec{\phi}_{\pfromq}^m,1) \bigm]. \nonumber
\evq
}
 Working on the single expectations leads to
{\small
\bvq
 \mathcal{L}_\Delta (q,\Theta)
 &=& \sum_{m} \sum_{p,q} \sum_{g,h}  \phi_{\ptoq, g}^m \phi_{\pfromq, h}^m \cdot f \bigm( R_m(p,q), B(g,h) \bigm) \nonumber \\
 &+& \sum_m \sum_{p,q} \sum_{g} \phi_{\ptoq, g}^m \bigm[ \psi(\gamma_{p,g}) - \psi(\sum_{g} \gamma_{p,g})\bigm] \nonumber \\
 &+& \sum_m \sum_{p,q} \sum_{h} \phi_{\pfromq, h}^m \bigm[ \psi(\gamma_{p,h}) - \psi(\sum_{h} \gamma_{p,h})\bigm] \nonumber \\
 &+& \sum_{p} \log \Gamma(\sum_{k} \alpha_k) - \sum_{p,k} \log \Gamma(\alpha_k) 
 + \sum_{p,k} (\alpha_k -1) \bigm[ \psi(\gamma_{p,k}) - \psi(\sum_{k} \gamma_{p,k})\bigm] \nonumber \\
 &-& \sum_p \log \Gamma(\sum_k \gamma_{p,k}) + \sum_{p,k} \log \Gamma(\gamma_{p,k}) 
 - \sum_{p,k} (\gamma_{p,k} -1) \bigm[ \psi(\gamma_{p,k}) - \psi(\sum_k \gamma_{p,k})\bigm] \nonumber \\
 &-& \sum_m \sum_{p,q} \sum_g \phi_{\ptoq, g}^m \log \phi_{\ptoq, g}^m 
 - \sum_m \sum_{p,q} \sum_h \phi_{\pfromq, h}^m \log \phi_{\pfromq, h}^m \nonumber
\evq
}
%
%
 where
\[
 f \bigm( R_m(p,q), B(g,h) \bigm)  = R_m(p,q) \log B(g,h) + \bigm(1-R_m(p,q) \bigm) \log \bigm(1-B(g,h) \bigm);
\]
%
%
%
%
%
%
%
 $m$ runs over $1,\dots,M$; $p,q$ run over $1,\dots,N$; $g,h,k$ run over $1,\dots,K$; and $\psi (x)$ is the derivative of the log-gamma function, $\frac{d\log \Gamma(x)}{dx}$.

\subsection{The Expected Value of the Log of a Dirichlet Random Vector}
\label{app:exp_log_pi}

 The computation of the lower bound for the likelihood requires us to evaluate $ \mathbb{E}_q \bigm[\log \vec{\pi}_p\bigm]$ for $p = 1,\dots,N$.
 Recall that the density of an exponential family distribution with natural parameter $\vec{\theta}$ can be written as
\bvq
 p (x|\alpha)
 & = & h(x) \cdot c(\alpha) \cdot \exp \bigm\{ \sum_k ~ \theta_k(\alpha) \cdot t_k(x) \bigm\} \nonumber \\
 & = & h(x) \cdot \exp \bigm\{ \sum_k ~ \theta_k(\alpha) \cdot t_k(x) - \log c(\alpha) \bigm\}. \nonumber
\evq
 Omitting the node index $p$ for convenience, we can rewrite the density of the Dirichlet distribution $p_3$ as an exponential family distribution,
\[
 p_3 (\vec{\pi} | \vec{\alpha}) = \exp \biggm\{ \sum_k (\alpha_k-1) \log (\pi_{k}) - \log \frac{\prod_k \Gamma (\alpha_k)} { \Gamma (\sum_k \alpha_k )} \biggm\},
\]
 with natural parameters $\theta_k (\vec{\alpha}) = (\alpha_k-1)$ and natural sufficient statistics $t_k(\vec{\pi}) = \log(\pi_{k})$.
 Let $c'(\vec{\theta}) = c (\alpha_1(\vec{\theta}), \dots, \alpha_K(\vec{\theta}))$; using a well known property of the exponential family distributions \cite{Sche:1995} we find that
\bvq
 \mathbb{E}_q \bigm[ \log \pi_{k} \bigm]
 & = & \mathbb{E}_{\vec{\theta}} \bigm[ \log t_k (x) \bigm] 
 = \psi \bigm(\alpha_k \bigm) - \psi \bigm(\sum_k \alpha_k \bigm), \nonumber
\evq
 where $\psi (x)$ is the derivative of the log-gamma function, $\frac{d\log \Gamma(x)}{dx}$.

\subsection{Variational E Step}
\label{sec:inf_e}

 The approximate lower bound for the likelihood $\mathcal{L}_\Delta (q,\Theta)$ can be maximized using exponential family arguments and coordinate ascent \cite{Wain:Jord:2003b}.

 Isolating terms containing $\phi_{\ptoq,g}^m$ and $\phi_{\pfromq,h}^m$ we obtain $\mathcal{L}_{\phi_{\ptoq,g}^m} (q,\Theta)$ and $\mathcal{L}_{\phi_{\ptoq,g}^m} (q,\Theta)$.
%
%
%
%
 The natural parameters $\vec{g}_{\ptoq}^m$ and $\vec{g}_{\pfromq}^m$ corresponding to the natural sufficient statistics $\log(\vec{z}_{\ptoq}^m)$ and $\log(\vec{z}_{\pfromq}^m)$ are functions of the other latent variables and the observations.  We find that
\bvq
 g_{\ptoq,g}^m
 & = & \log \pi_{p,g} + \sum_{h} z_{\pfromq,h}^m \cdot f \bigm( R_m(p,q), B(g,h) \bigm), \nonumber  \\
 g_{\pfromq,h}^m
 & = & \log \pi _{q,h} + \sum_{g} z_{\ptoq,g}^m \cdot f \bigm( R_m(p,q), B(g,h) \bigm), \nonumber
\evq
 for all pairs of nodes $(p,q)$ in the $m$-$th$ network; where $g,h=1, \dots, K$, and
\[
 f \bigm( R_m(p,q), B(g,h) \bigm) = R_m(p,q) \log B(g,h) + \bigm(1-R_m(p,q) \bigm) \log \bigm(1-B(g,h) \bigm).
\]
 This leads to the following updates for the variational parameters $(\vec{\phi}_{\ptoq}^m,\vec{\phi}_{\pfromq}^m)$, for a pair of nodes $(p,q)$ in the $m$-$th$ network:
\bvq
 \label{eq:phi_to}
 \hat \phi_{{\ptoq, g}}^m
 & \propto &
 e^{~\mathbb{E}_q \bigm[ g_{\ptoq,g}^m \bigm]} \\
 & = &
 e^{~\mathbb{E}_q \bigm[ \log \pi_{p,g} \bigm]} \cdot e^{~\sum_h \phi_{\pfromq,h}^m \cdot ~\mathbb{E}_q \bigm[ f \bigm( R_m(p,q), B(g,h) \bigm) \bigm]} \nonumber \\
 & = &
 e^{~\mathbb{E}_q \bigm[ \log \pi_{p,g} \bigm]} \cdot \prod_h \biggm( B(g,h)^{R_m(p,q)} \cdot \bigm(1-B(g,h) \bigm)^{1-R_m(p,q)} \biggm)^{\phi_{\pfromq,h}^m}, \nonumber \\
 \label{eq:phi_from}
 \hat \phi_{{\pfromq, h}}^m
 & \propto &
 e^{~\mathbb{E}_q \bigm[ g_{\pfromq,h}^m \bigm]} \\
 & = &
 e^{~\mathbb{E}_q \bigm[ \log \pi_{q,h} \bigm]} \cdot e^{~\sum_g \phi_{\ptoq,g}^m \cdot ~\mathbb{E}_q \bigm[ f \bigm( R_m(p,q), B(g,h) \bigm) \bigm]} \nonumber \\
 & = &
 e^{~\mathbb{E}_q \bigm[ \log \pi_{q,h} \bigm]} \cdot \prod_g \biggm( B(g,h)^{R_m(p,q)} \cdot \bigm(1-B(g,h) \bigm)^{1-R_m(p,q)} \biggm)^{\phi_{\ptoq,g}^m}, \nonumber
\evq
 for $g,h=1, \dots, K$. These estimates of the parameters underlying the distribution of the nodes' group indicators $\vec{\phi}_{\ptoq}^m$ and $\vec{\phi}_{\pfromq}^m$ need be normalized, to make sure $\sum_k \phi_{\ptoq,k}^m = \sum_k \phi_{\pfromq,k}^m = 1$.

 Isolating terms containing $\gamma_{p,k}$ we obtain $\mathcal{L}_{\gamma_{p,k}} (q,\Theta)$.
%
 Setting $\frac{\partial \mathcal{L}_{\gamma_{p,k}}}{\partial \gamma_{p,k}}$ equal to zero and solving for $\gamma_{p,k}$ yields:
\bvq
 \label{eq:gamma}
 \hat \gamma_{p,k} = \alpha_k + \sum_m \sum_{q} \phi_{\ptoq,k}^m + \sum_m \sum_{q} \phi_{\pfromq,k}^m,
\evq
 for all nodes $p\in\mathcal{P}$ and $k=1, \dots, K$.

 The $t$-$th$ iteration of the variational E step is carried out for fixed values of $\Theta^{(t-1)} = (\vec{\alpha}^{(t-1)},B^{(t-1)})$, and finds the optimal approximate lower bound for the likelihood $\mathcal{L}_{\Delta^*} (q,\Theta^{(t-1)})$.

\subsection{Variational M Step}
\label{sec:inf_m}

 The optimal lower bound $\mathcal{L}_{\Delta^*} (q^{(t-1)},\Theta)$ provides a tractable surrogate for the likelihood at the $t$-$th$ iteration of the variational M step. We derive empirical Bayes estimates for the hyper-parameters $\Theta$ that are based upon it.\footnote{We could term these estimates {\it pseudo} empirical Bayes estimates, since they maximize an approximate lower bound for the likelihood, $\mathcal{L}_{\Delta^*}$.}
%
%
 That is, we maximize $\mathcal{L}_{\Delta^*} (q^{(t-1)},\Theta)$ with respect to $\Theta$, given expected sufficient statistics computed using $\mathcal{L}_{\Delta^*} (q^{(t-1)},\Theta^{(t-1)})$.

 Isolating terms containing $\vec{\alpha}$ we obtain $\mathcal{L}_{\vec{\alpha}}(q,\Theta)$. Unfortunately,
%
%
 a closed form solution for the approximate maximum likelihood estimate of $\vec\alpha$ does not exist \cite{Blei:Ng:Jord:2003}.  We can produce a Newton-Raphson method that is linear in time, where the gradient and Hessian for the bound $\mathcal{L}_{\vec{\alpha}}$ are
\begin{eqnarray*}
   \frac{\partial \mathcal{L}_{\vec{\alpha}}}{\partial \alpha_{k}} & = &
    N \biggm( \psi \bigm( \sum_{k} \alpha_{k} \bigm) - \psi (\alpha_{k}) \biggm) + \sum_{p} \biggm( \psi(\gamma_{p,k}) - \psi \bigm( \sum_{k} \gamma_{p,k} \bigm) \biggm),\\
   \frac{\partial \mathcal{L}_{\vec{\alpha}}}{\partial \alpha_{k_1} \alpha_{k_2}} & = &
    N \biggm( \mathbb{I}_{(k_1 = k_2)} \cdot \psi' (\alpha_{k_1}) - \psi' \bigm( \sum_{k} \alpha_{k} \bigm) \biggm).
\end{eqnarray*}

 Isolating terms containing $B$ we obtain $\mathcal{L}_{B}$,  whose approximate maximum is
%
%
%
\bvq
 \label{eq:b}
 \hat B (g,h) = \frac{1}{M} \sum_m \biggm( \frac{\sum_{p,q} R_m(p,q) \cdot \phi_{\ptoq g}^m \, \phi_{\pfromq h}^m}{\sum_{p,q} \phi_{\ptoq g}^m \, \phi_{\pfromq h}^m}\biggm),
\evq
 for every index pair $(g,h) \in [1,K] \times [1,K]$.

 In Section \ref{sec:sparsity} we introduced an extra parameter, $\rho$, to control the relative importance of presence and absence of interactions in likelihood, i.e., the score that informs inference and estimation.
 Isolating terms containing $\rho$ we obtain $\mathcal{L}_{\rho}$.
%
%
 We may then estimate the sparsity parameter $\rho$ by
\bvq
 \label{eq:rho}
 \hat \rho = \frac{1}{M} \sum_m \biggm( \frac{\sum_{p,q} \bigm( 1-R_m(p,q) \bigm) \cdot \bigm( \sum_{g,h} \phi_{\ptoq g}^m \, \phi_{\pfromq h}^m\bigm)}{\sum_{p,q} \sum_{g,h} \phi_{\ptoq g}^m \, \phi_{\pfromq h}^m} \biggm).
\evq
 Alternatively, we can fix  $\rho$ prior to the analysis; the density of the interaction matrix is estimated with $\hat d = \sum_{m,p,q} R_m(p,q)/(N^2M)$, and the sparsity parameter is set to $\tilde \rho = (1-\hat d)$.  This latter estimator attributes all the information in the non-interactions to the point mass, i.e., to latent sources other than the block model $B$ or the mixed membership vectors $\vec{\pi}_{1:N}$.  It does, however, provide a quick recipe to reduce the computational burden during exploratory analyses.\footnote{Note that $\tilde \rho = \hat \rho$ in the case of single membership. In fact, that implies $\phi_{\ptoq g}^m = \phi_{\pfromq h}^m =1$ for some $(g,h)$ pair, for any $(p,q)$ pair.}

\end{document}